\documentclass[]{aa}

\newcommand*{\crosssymbol}{%
    \text{%
      \raisebox{1ex}{%
        \makebox[0pt][l]{%
          \rule[-.2pt]{.75ex}{.4pt}%
        }%
        \makebox[.75ex]{%
          \rule[-1ex]{.4pt}{1.5ex}%
        }%
      }%
    }%
}   
\usepackage{xtab,booktabs}
\usepackage{ltablex}
\usepackage{changepage}
\usepackage{graphicx,lscape}
\usepackage{subfig}
\usepackage{txfonts,longtable}
\usepackage{graphicx,rotating}
\usepackage{subfig}
\usepackage{txfonts}
\usepackage[T1]{fontenc}
\usepackage{epsfig}
\usepackage{natbib}
\usepackage{color}
\usepackage{natbib}
\usepackage{amssymb}  
\usepackage{tabularx}
\usepackage{comment}
\usepackage{stackengine}
\def\cross{%
  \stackon[1ex]{\rule{0.4pt}{1.5ex}}{\rule{.75ex}{0.4pt}}}

\DeclareGraphicsExtensions{.pdf,.png,.jpg,.ps}
\bibpunct{(}{)}{;}{a}{}{,} 
\begin{document}

   \title{Molecular gas in CLASH brightest cluster galaxies at $z\sim0.2-0.9$}

   \author{G. Castignani
          \inst{1}\fnmsep\thanks{e-mail: gianluca.castignani@epfl.ch}
          \and
          M.~Pandey-Pommier\inst{2,3}
          \and
          S.~L. Hamer\inst{4}
          \and
          F. Combes\inst{5,6}
          \and
          P. Salom\'e\inst{5}
          \and
          J. Freundlich\inst{7,8}
          \and
          P. Jablonka\inst{1,9}
          }

   \institute{Laboratoire d'astrophysique, \'{E}cole Polytechnique F\'{e}d\'{e}rale de Lausanne, Observatoire de Sauverny, 1290 Versoix, Switzerland
              \and
             USN, Station de Radioastronomie de Nan\c{c}ay, Observatoire de Paris, route de Souesmes, 18330 Nan\c{c}ay, France
             \and
             Univ. Lyon, Ens de Lyon, CNRS, Centre de Recherche Astrophysique   de   Lyon,   UMR5574,   69230   Saint-Genis-Laval, France
             \and
             Department of Physics, University of Bath, Claverton Down, BA2 7AY, UK
            \and
         Observatoire de Paris, LERMA, CNRS, Sorbonn\'{e} University, PSL Research Universty, 75014 Paris, France 
         \and
             Coll\`{e}ge de France, 11 Place Marcelin Berthelot, 75231 Paris, France
            \and
           Centre for Astrophysics and Planetary Science, Racah Institute of Physics, The Hebrew University, Jerusalem 91904, Israel
           \and
           School of Physics and Astronomy, Tel Aviv University, Tel Aviv 69978, Israel
           \and
          Observatoire de Paris, GEPI, CNRS, Sorbonn\'{e} University, PSL Research Universty, 75014 Paris, France
              }
                            \date{Received: April 3, 2020; Accepted June 2, 2020}

  \abstract
   {Brightest cluster galaxies (BCGs) are excellent laboratories for the study of galaxy evolution in dense Mpc-scale environments. We used the IRAM-30m to observe, in CO(1$\rightarrow$0), CO(2$\rightarrow$1), CO(3$\rightarrow$2), or CO(4$\rightarrow$3), 18 BCGs at $z\sim0.2-0.9$   drawn from the Cluster Lensing And Supernova survey with {\it Hubble} (CLASH) survey. Our sample includes RX1532, which is  our primary target as it is among the BCGs with the highest star formation rate (SFR$\gtrsim100~M_\odot$/yr) in the CLASH sample. We unambiguously detected both CO(1$\rightarrow$0) and CO(3$\rightarrow$2) in RX1532, yielding a large reservoir of molecular gas,  $M_{H_2}=(8.7\pm1.1)\times10^{10}~M_\odot$, and a high level of excitation, $r_{31}=0.75\pm0.12$. A morphological analysis of the {\it Hubble Space Telescope} I-band image of RX1532 reveals the presence of clumpy substructures both within and outside the half-light radius $r_e=(11.6\pm0.3)$~kpc, similarly to those found independently both in ultraviolet and in H$_\alpha$ in previous works. We tentatively detected CO(1$\rightarrow$0) or CO(2$\rightarrow$1) in four other BCGs, with molecular gas reservoirs in the range of $M_{H_2}=2\times10^{10-11}~M_\odot$. For the remaining 13 BCGs,  we set robust upper limits of $M_{H_2}/M_\star\lesssim0.1$, which are
   among the lowest molecular-gas-to-stellar-mass ratios found for distant ellipticals and BCGs. In  comparison with distant cluster galaxies observed in CO, our study shows that RX1532 ($M_{H_2}/M_\star = 0.40\pm0.05$) belongs to the rare population of star-forming and gas-rich BCGs in the distant universe. By using the available X-ray based estimates of the central intra-cluster medium entropy, we show that the detection of large reservoirs of molecular gas $M_{H_2}\gtrsim10^{10}~M_\odot$ in distant BCGs is possible when the two conditions are met: i) high SFR and ii) low central entropy, which favors the condensation and the inflow of gas onto
   the BCGs themselves, similarly to what has been previously found for some local BCGs.}

   


   \maketitle
%

\section{Introduction}\label{sec:introduction}
Brightest Cluster Galaxies (BCGs) are among the most massive and luminous galaxies in the Universe. They are almost invariably associated with passively evolving massive ellipticals of cD type and with radio galaxies, at least locally \citep{Zirbel1996}.
BCGs are also located at the centers of galaxy clusters  \citep{Lauer2014}, where high intra-cluster medium (ICM) densities, high galaxy number densities, and large relative velocities between cluster members occur. BCGs are, therefore, optimal laboratories for the study of the co-evolution and interplay of (active) galaxies with their Mpc-scale environments in extreme conditions.
Indeed, BCGs are believed to evolve via environmental processes, such as dynamical friction \citep{White1976}, galactic cannibalism \citep{Hausman_Ostriker1978}, interactions with the intracluster medium \citep{Stott2012}, and cooling flows \citep{Salome2006}. However, the relative weight of these processes, responsible for the stellar mass assembly of BCGs is still under debate.

Significant efforts have been carried out to investigate the star formation history of BCGs in order 
to better understand their evolution and growth.
Previous work found significant ultraviolet (UV) and far-infrared (FIR) emission in both local and more distant BCGs \citep{Hicks2010,Rawle2012,Fogarty2015,Webb2015,Bonaventura2017}. Although active galactic nucleus (AGN) activity may contaminate the UV and FIR continuum of BCGs,  some of them appear to be  unambiguously characterized by high star-formation activity, such as the well-known Perseus~A and Cygnus~A galaxies, with their star formation rates (SFR) of $>40~M_\odot$/yr  \citep{FraserMcKelvie2014}.

Other studies of BCGs, mostly local, have detected large molecular gas reservoirs  \citep{Edge2001,Salome_Combes2003,Hamer2012,McNamara2014, Russell2014,Tremblay2016} and even filaments of cold gas \citep{Olivares2019,Russell2019}.   
Some spectacular star-forming and gas-rich BCGs have also been observed in the distant Universe. \citet{McDonald2014} found a large molecular gas reservoir $M_{H_2}\simeq2\times10^{10}~M_\odot$ in the Phoenix~A BCG at $z=0.597$. This finding implies a rapid stellar mass growth for the BCG, since a significant fraction  $f\simeq3\%$ of the BCG stellar mass is accumulated within its starburst phase  \citep[SFR$\sim800~M_\odot/$yr,][]{McDonald2013} in a relatively short depletion timescale, $\tau_{\rm dep}\sim30$~Myr.

More recently, \citet{Fogarty2019} detected a similar amount of molecular gas
$M_{H_2}\gtrsim10^{10}~M_\odot$ in the star-forming BCG of the cool-core cluster MACS~1931.8-2635 (abbreviated as M1932 in this work). Emission from several CO lines have been found by the authors, associated with both compact and more diffuse components. The latter remarkably traces a knotty filament revealed both in UV  and H$\alpha$ by the {\it Hubble Space Telescope (HST)}, out to  $\sim$30~kpc to the northwest of the BCG. 

AGN feedback-regulated cooling in the ICM is a commonly invoked mechanism to explain the fact that large reservoirs of molecular gas and high star formation activity are rarely observed in the BCGs: condensation of the ICM occurs in the central and low-entropy regions of the cluster, while the energy injected into the ICM via radio jets propagating from the BCG cores themselves prevents the formation of  $\sim(100{\rm s}-1000{\rm s})~M_\odot$/yr cooling flows that were theoretically  predicted  \citep[e.g.,][]{Fabian1994,McNamara2000,Peterson2003,Peterson_Fabian2006,McNamara_Nulsen2012}.

In the present study, we investigate the molecular gas properties and their evolution for a well-studied sample of BCGs at $z\sim0.2-0.9$ using new IRAM 30m observations in combination with ancillary information. The sources are drawn from a sub-sample of clusters from the  Cluster Lensing And Supernova survey with {\it Hubble} (CLASH) survey. CLASH clusters have high-resolution, multi-band {\it HST} observations \citep{Postman2012a} and deep X-ray monitoring with {\it Chandra} \citep{Donahue2016}. Several spectroscopic campaigns also led to the confirmation of a large number of CLASH cluster members \citep[][and references therein]{Caminha2019}. CLASH is therefore an excellent survey to investigate galaxy evolution in clusters \citep{Annunziatella2014,Annunziatella2016} and to study both BCGs independently \citep{Postman2012b,Burke2015,Yu2018,Durret2019} and while factoring in   their interplay with the ICM \citep{Donahue2016,Fogarty2017,DeMaio2019}.



This work is part of a larger search done over a wide range of redshifts, $z\sim0.4-2.6$, to detect molecular gas in distant BCGs \citep{Castignani2019,Castignani2020a,Castignani2020b}.
The paper is  structured as follows: in Sect.~\ref{sec:sample} we describe the sample of target BCGs; in Sect.~\ref{sec:morph_RX1532} we present the morphological analysis for the BCG RX1532 in our sample; in Sect.~\ref{sec:observations_and_data_reduction} we describe the observations and data reduction; in Sects.~\ref{sec:results} and \ref{sec:discussion}  we present and discuss the results; in Sect.~\ref{sec:conclusions} we draw our conclusions. 

{ For the CLASH BCGs considered in this work, stellar mass and star formation rate estimates come from previous studies \citep{Burke2015,Fogarty2015,Donahue2015} and rely on the \citet{Salpeter1955} initial mass function (IMF). The only exception is the stellar mass estimate for the M1932 BCG \citep{Cooke2016}, which relies on the \citet{Chabrier2003} IMF.
The use of a different IMF for M1932 does not significantly affect our results. 
We refer to \citet{Cooke2016} and Sect.~\ref{sec:RX1532} for further discussion.} 
Throughout this work we adopt a flat $\Lambda \rm CDM$ cosmology with matter density $\Omega_{\rm m} = 0.30$, dark energy density $\Omega_{\Lambda} = 0.70$, and Hubble constant $h=H_0/(100\, \rm km\,s^{-1}\,Mpc^{-1}) = 0.70$ \citep[but see][]{PlanckCollaborationVI2018,Riess2019b}.

\begin{table*}[]\centering
\begin{adjustwidth}{-0.5cm}{}
\begin{center}
\begin{tabular}{ccccccccc}
\hline\hline
Galaxy  & R.A. & Dec. & $z_{spec}$ &  $M_\star$ & SFR & sSFR & sSFR$_{\rm MS}$ & $K_0$  \\
   & (hh:mm:ss.s) & (dd:mm:ss.s) &  &  ($10^{11}~M_\odot$) & ($M_\odot$/yr) & (Gyr$^{-1}$) & (Gyr$^{-1}$) & (keV~cm$^2$) \\ 
  (1) & (2) & (3) & (4) & (5) & (6) & (7) & (8) & (9) \\
 \hline
A209  &     01:31:52.53 &  -13:36:40.5   & 0.206 & $2.00\pm0.16$ & $<0.12$~(UV) & $<6.0\times10^{-4}$ & 0.04 & $106\pm27$\\ 
\hline
M0329 &     03:29:41.57 &  -02:11:46.6    & 0.450 & $3.69\pm0.21$ & $42\pm2$~(UV) & $0.11\pm0.01$ & 0.07 & $11\pm3$\\ 
  &             &               &         &                 & $80\pm21$~(H$\alpha$) & $0.22\pm0.06$  &  &\\
  \hline
M0429 &        04:29:36.03 &  -02:53:06.8 & 0.399 & $4.68\pm0.23$ & $28\pm2$~(UV) & $0.06\pm0.01$ & 0.06 & $17\pm4$ \\ 
  &             &               &         &                 & $33\pm9$~(H$\alpha$) & $0.07\pm0.02$  &  &\\
  \hline
M0647 &     06:47:50.70  &   +70:14:54.0  & 0.591 & $5.13\pm1.79$ & $<0.3$~(UV) & $<5.8\times10^{-4}$ & 0.10 & $225\pm50$ \\ 
\hline
M0744       &  07:44:52.80 &  +39:27:26.5 & 0.686 & $8.20\pm0.43$ & $0.6\pm0.1$~(UV) & $(7.3\pm1.3)\times10^{-4}$ & 0.11 & $42\pm11$ \\ 
\hline
A611        &  08:00:56.82 &  +36:03:23.6 & 0.288 & $3.45\pm0.42$ & $<0.04$~(UV)& $<1.2\times10^{-4}$ & 0.04 & $125\pm18$ \\ 
\hline
M1115       &  11:15:51.90 &  +01:29:55.0 & 0.352 & $2.19\pm0.17$ & $13\pm1$~(UV) &
$(5.5\pm0.6)\times10^{-2}$ & 0.07 & $15\pm3$ \\ 
  &             &               &         &                 & $3.4\pm1.0$~(H$\alpha$) & $(1.6\pm0.5)\times10^{-2}$   &  &\\
\hline
A1423       &   11:57:17.36 &   +33:36:39.7 & 0.213 & $1.96\pm0.11$ & $0.07\pm0.03$~(UV)& $(3.6\pm1.5)\times10^{-4}$ & 0.04 & $68\pm13$ \\ 
\hline
M1206       &  12:06:12.17 &  -08:48:02.9  & 0.440 & $4.93\pm1.39$ & $2.9\pm0.6$~(UV)& $(5.9\pm2.1)\times10^{-3}$ & 0.06 & $69\pm10$ \\ 
\hline
CL1226 &  12:26:58.25 &   +33:32:48.6    & 0.891  & $15.00\pm1.99$   & $0.9\pm0.1$~(UV) & $(6.0\pm1.0)\times10^{-4}$  & 0.14 & $166\pm45$ \\ 
\hline
M1311       &  13:11:01.81 & -03:10:39.7 & 0.494 & $4.49\pm1.05$ & $<1.2$~(UV) & $<2.7\times10^{-3}$ & 0.08 & $47\pm4$ \\ 
\hline
RX1347      &  13:47:31.83 &  -11:45:11.6 & 0.451 & $3.68\pm0.85$ & $22\pm1$~(UV) & $(6.0\pm1.4)\times10^{-2}$ & 0.08 & $12\pm20$ \\ 
&             &               &         &                 & $12\pm4$~(H$\alpha$) & $(3.3\pm1.3)\times10^{-2}$   &  &\\
\hline  
M1423       &  14:23:47.88 &  +24:04:42.4 & 0.545 & $4.95\pm0.13$ & $27\pm1$~(UV) & $(5.5\pm0.2)\times10^{-2}$ & 0.09 & $10\pm5$ \\ 
&             &               &         &                 & $33\pm9$~(H$\alpha$) & $(6.7\pm1.9)\times10^{-2}$  &  &\\
\hline
 RX1532 & 15:32:53.79 & +30:20:59.5    & 0.361 &  $2.20\pm0.05$  & $97\pm4$~(UV) & 0.44$\pm$0.02  & 0.07  & $17\pm2$\\
        &             &               &         &                 & $140\pm40$~(H$\alpha$) & 0.64$\pm$0.18 &  & \\

 \hline
M1720 &   17:20:16.75 &  +35:36:26.2    & 0.391 &  $3.83\pm0.94$  & $1.1\pm0.2$~(UV) & $(2.9\pm0.9)\times10^{-3}$  & 0.06 & $24\pm3$\\ 
  &             &               &         &                 & $4.4\pm1.2$~(H$\alpha$) & $(1.1\pm0.4)\times10^{-2}$  &  &\\
  \hline
A2261 &   17:22:27.20 &  +32:07:56.9    & 0.224 &  $1.74\pm0.18$  & $<0.06$~(UV) & $<3.4\times10^{-4}$ & 0.04 & $61\pm8$\\ 
 \hline
M1932      &     19:31:49.6 &  -26:34:33.2 & 0.353 & $6.9\pm0.8$          &  $280\pm20$~(UV)           &   $0.41\pm0.06$    & 0.04   &  $14\pm4$   \\ 
     &     &  &  &           &  $130\pm40$~(H$\alpha$)           &   $0.19\pm0.06$    &        &  \\
\hline
M2129       &   21:29:26.12 &  -07:41:27.9 & 0.570 & $2.44\pm0.17$ & $<0.1$~(UV) & $<4.1\times10^{-4}$ & 0.13 & $200\pm100$ \\ 
\hline
RX2129      &  21:29:39.96 &  +00:05:21.1 & 0.234 & $2.21\pm0.32$ & $<3$~(UV) & $<0.01$ & 0.04 & $21\pm4$  \\
&             &               &         &                 & $<49$~(H$\alpha$) &  $<0.22$ &  &\\ 
\hline  
\end{tabular}
 \end{center}
\caption{Properties of our targets. (1) BCG name; (2-3) J2000 equatorial coordinates; (4) spectroscopic redshift; (5) stellar mass from \citet{Burke2015} except for M1932 \citep{Cooke2016} ; (6) UV- and H$\alpha$-based SFR, corrected for dust extinction, from \citet{Fogarty2015}, except for M0647 and M2129, for which the SFR upper limits are from \citet{Donahue2015}; (7) specific SFR determined as ${\rm sSFR}={\rm SFR}/M_\star$; (8) sSFR for main sequence field galaxies with redshift and stellar mass of our targets estimated using the relation found by \citet{Speagle2014}; (9) central entropy $K_0=k_B~T_X~n_e^{-2/3}$ of the cluster estimated by \citet{Donahue2015}. Upper limits are at $3\sigma$.}
\label{tab:BCG_properties}
\end{adjustwidth}
\end{table*}

\section{Sample of CLASH brightest cluster galaxies}\label{sec:sample}

As we aim to target actively star-forming BCGs in order to investigate their molecular gas content and explore their evolution and interaction with their Mpc-scale environments, we consider a homogeneous sample of 19 BCGs at $z\sim0.2-0.9$, drawn from the list of 25 CLASH clusters, as described in the following. Since our observational strategy primarily uses the IRAM 30m telescope in the northern hemisphere, for our selection we rejected the CLASH clusters MS~2137-2353 ($z=0.313$),  RX~J2248.7-4431 (i.e., Abell~1063S at $z=0.348$), and MACS~J0416.1-2403 ($z=0.42$), which are located at low declination Dec.$<-23$~degree, and are therefore difficult or impossible to observe with the IRAM 30m, at an elevation $>20$~degree.  While we should formally discard the  CLASH BCG M1932, at Dec.=-26:34:34.0, we included it in our BCG sample since it was recently detected in CO by \citet{Fogarty2019} and has a number of ancillary information from the literature, including a high star formation rate  SFR$\simeq(100-300)~M_\odot$/yr \citep{Fogarty2015} and a  stellar mass estimate $M_\star\simeq7\times10^{11}~M_\odot$ \citep{Cooke2016}. As part of our IRAM 30m campaign we observed all remaining CLASH BCGs, at declination Dec.$>$-14~degree, with the exception of the BCGs of Abell~383 ($z=0.187$), MACS~J1149.5+2223 ($z=0.544$), and MACS~J0717.5+3745 ($z=0.546$). Concerning this last cluster, the molecular gas properties of a sub-population of star-forming galaxies are investigated in a separate work \citep{Castignani2020c}. 

The majority, that is, 12 out of the 19 BCGs, have clear UV features identified by \citet{Fogarty2015} using {\it HST} maps, while 8 among the 12 also have H$\alpha$+[NII] emission detected via CLASH {\it HST} data. The combined use of the UV continuum and H$\alpha$+[NII] maps allowed the authors to derive robust SFR estimates for the great majority of the BCGs in our sample, which range from low values of $\lesssim1~M_\odot$/yr to very high values of $\gtrsim100~M_\odot$/yr as in the case of RX1532 and M1932. The CLASH BCGs of our sample are also hosted in the cores of clusters with different dynamical states, as quantified from the cluster core entropy $K_0=k_B~T_X~n_e^{-2/3}\simeq(10-200)$~keV~cm$^2$ \citep[see e.g.,][for a review]{Voit2005}, estimated from X-ray analysis by \citet{Donahue2014,Donahue2015} using {\it Newton-XMM} and {\it Chandra} observations of CLASH clusters.
Some properties of the targeted sources are listed in Table~\ref{tab:BCG_properties}, including stellar masses from \citet{Burke2015}, which were carefully estimated by  the authors using CLASH multiband {\it HST} photometry and taking a possible contamination due to the ICM and nearby companions of the BCGs into account.  

\section{Morphological analysis of RX1532}\label{sec:morph_RX1532}
In this section, we investigate the morphology of the RX1532 BCG at $z=0.36$. This BCG is among those of the \citet{Fogarty2015} sample which have the highest SFR$\simeq100~M_\odot$/yr, together with MACS~1931.8-2635 BCG. Both BCGs are also hosted in dynamically relaxed cool-core clusters with entropy floors $K_0\lesssim20$~keV~cm$^2$, which are among the lowest found within the CLASH sample  \citep{Donahue2015,Donahue2016}. 

\begin{figure*}[]\centering
\captionsetup[subfigure]{labelformat=empty}
\subfloat[]{\hspace{0.cm}\includegraphics[trim={0cm 0cm 0cm 
0cm},clip,width=0.25\textwidth,clip=true]{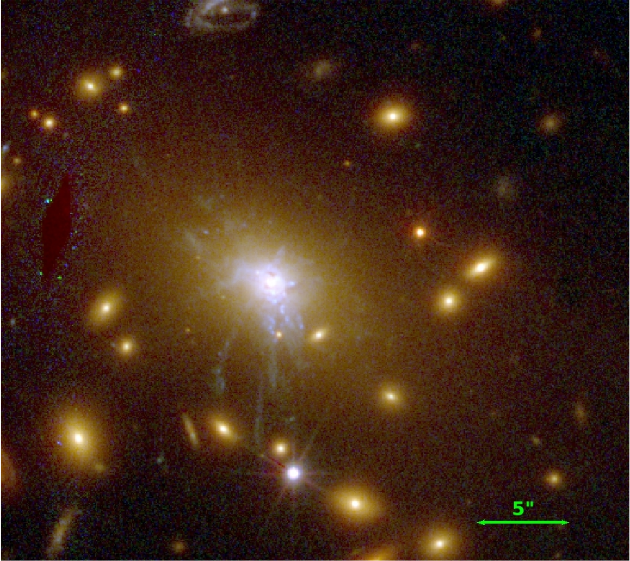}}
\subfloat[]{\hspace{0.7cm}\includegraphics[trim={0cm 0cm 0cm 
0cm},clip,width=0.35\textwidth,clip=true]{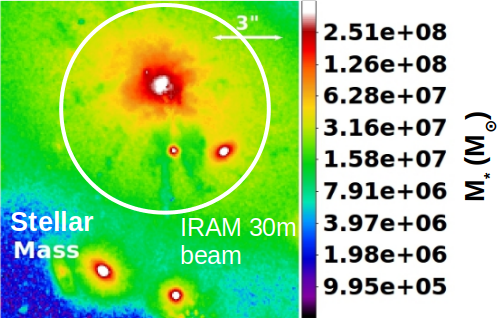}}
\subfloat[]{\hspace{0.7cm}\includegraphics[trim={0cm 0cm 0cm 
0cm},clip,width=0.3\textwidth,clip=true]{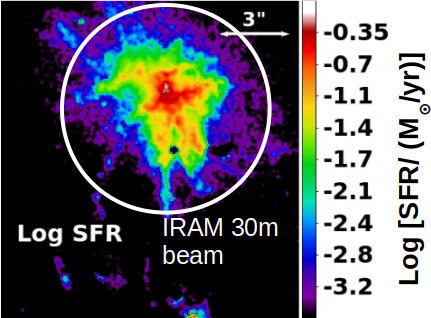}}
\caption{Color composite {\it HST} image (left), stellar mass (center), and SFR (right) density maps of RX1532 ($z=0.36$) taken from \citet{Fogarty2015}. In the left panel the RGB color composite is made using WFC3-IR filters {\it F105W+F110W+F125W+F140W+F160W} in red, the ACS filters {\it F606W+F625W+F775W+F814W+F850LP} in green, and the ACS filters {\it F435W+F475W} in blue. In the center and left panels, we overplot the IRAM 30m beam of our CO(3$\rightarrow$2) observations.}\label{fig:RX1532_images}
\end{figure*}

In Fig.~\ref{fig:RX1532_images}, we show the {\it HST} image and both the stellar mass and SFR density maps of RX1532, taken from \citet{Fogarty2015}.  The figure shows a complex morphology both in stellar  mass and in star-formation for RX1532, with a tentative evidence for star-forming filaments extending from the BCG in several directions, similarly to the MACS~1931.8-2635 BCG.

\begin{figure*}[htb]\centering
\captionsetup[subfigure]{labelformat=empty}
\subfloat[]{\hspace{0.cm}\includegraphics[trim={0cm 0cm 0cm 
0cm},clip,width=1.0\textwidth,clip=true]{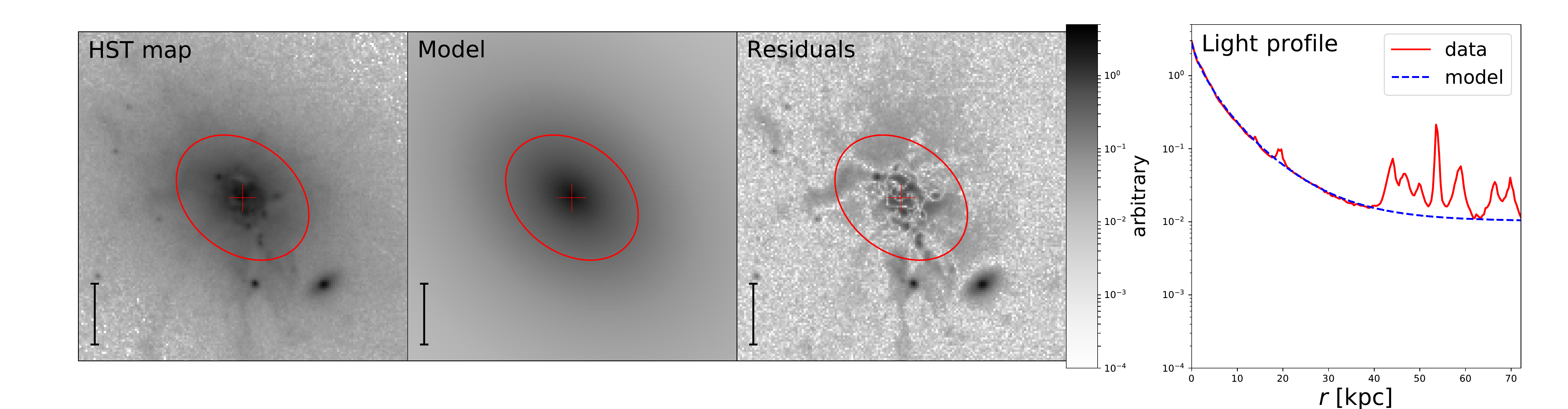}}
\caption{Morphological analysis of RX1532. From left to right: {\it HST} ACS image (F814W filter), best-fit single S\'{e}rsic model, residuals, and averaged radial light profile with the associated model. The red ellipse and the red cross in the images show the location of the half-light radius and that of the BCG, respectively. The vertical segment at the bottom left of each image has a size of 2~arcsec (i.e.,10~kpc).}\label{fig:RX1532_galfit}
\end{figure*}

The complex morphology observed in RX1532 and its similarities with M1932, motivated us to further investigate it  using {\sc Galfit} \citep{Peng2002,Peng2010}. Here we note that \citet{Durret2019} already performed a morphological analysis of CLASH BCGs using {\sc Galfit}. However, they did not include RX1532 in their sample.  For our analysis, we used the {\it HST} ACS F814W (I-band) image of RX1532 and the associated point spread function, both publicly available in the CLASH archive\footnote{https://archive.stsci.edu/prepds/clash/}. The {\it HST} image has a high resolution of 65~mas, that is, 0.3~kpc at the redshift of the BCG. The fit was performed using a single S\`{e}rsic profile, as reported in Eq.~(3) of \citet{Castignani2019}. The results are reported in Fig.~\ref{fig:RX1532_galfit}.

Our best fit is reasonably good up to a radius of $\sim$40~kpc. It also yielded a half-light radius of $r_e=(11.6\pm0.3)$~kpc, which is { slightly higher than the values of $r_e\sim9.9$~kpc and $r_e\sim10.8$~kpc predicted  by \citet{vanderWel2014} for star-forming (late-type) and early-type galaxies, respectively, of similar mass and redshift than RX1532. }

{ Our {\sc Galfit} analysis also yielded a S\'{e}rsic index $n\simeq1.7$, which corresponds a less concentrated profile than those typical of elliptical galaxies ($n\sim4$).} Overall, our results imply that RX1532 is extended, as also suggested by the complex morphology observed by {\it HST}.
In fact, similarly to what has been observed in Fig.~\ref{fig:RX1532_images}, several clumpy substructures and  filamentary extended components are clearly visible from the map of the residuals in Fig.~\ref{fig:RX1532_galfit}, both within and outside the half-light radius, as denoted by the red ellipses.
These results strengthen the fact that several compact star-forming regions are indeed located in and around the BCG, up to $\sim${ 20}~kpc.

In the following section, we describe our IRAM-30m observations of the CLASH BCGs. The complex morphological properties and the exceptional star formation activity of RX1532, outlined above, strongly motivated us to consider it as our primary target.

\section{IRAM 30m observations and data reduction}\label{sec:observations_and_data_reduction}
We observed the CLASH BCGs in our sample, all except for M1932 \citep{Fogarty2019},  using the IRAM 30m telescope at Pico Veleta in Spain. The observations were carried out in July 2017 (ID: 065-17; PI: S.~Hamer), in May and August 2018 (ID:173-17, 088-18; PI: M.~Pandey-Pommier), and in January 2020 (ID: 209-19; PI: G.~Castignani).



We used the Eight Mixer Receiver (EMIR) to observe CO(J$\rightarrow$J-1) emission lines from the target sources, where J is a positive integer denoting the total angular momentum. For each source, the specific CO(J$\rightarrow$J-1) transitions were chosen to maximize the likelihood of the detection, in terms of the ratio of the predicted signal to the expected rms noise. We also preferred low-J transitions, that is, J~$=2$, 1, when possible.

The E090, E150, and E230 receivers, operating between $\sim$1-3~mm, offer 4$\times$4~GHz instantaneous bandwidth covered by the correlators. We used both E090 and E150 receivers, simultaneously, to observe CO(1$\rightarrow$0) and CO(2$\rightarrow$1) redshifted at $\sim$3~mm and $\sim$2~mm, respectively, when the atmosphere or the receiver ranges permitted it. CL1226 and RX1532 are the only exceptions, as described below. For CL1226 we targeted the CO(4$\rightarrow$3) line, redshifted at $\sim$1~mm, with the  E230 receiver. RX1532 was observed in CO(1$\rightarrow$0) with the E090 receiver in July 2017. In Jan 2020, we observed the source in  CO(1$\rightarrow$0) and CO(3$\rightarrow$2), simultaneously, using both receivers E090 and E230, respectively. { The IRAM 30m half power beam width is $\sim16$~arcsec~$\frac{\lambda_{\rm obs}}{2~{\rm mm}}$ \citep{Kramer2013}, where $\lambda_{\rm obs}$ is the observer frame wavelength. All targets were thus unresolved by our observations, as further discussed in Sect.~\ref{sec:results}.}

The wobbler-switching mode was used for all the observations to minimize the impact of atmospheric variability.
The Wideband Line Multiple Autocorrelator (WILMA) was used to cover the 4$\times$4~GHz bandwidth, in each linear polarization.   We also simultaneously recorded the data with the Fast Fourier Transform Spectrometers (FTS), as a backup, at 200 kHz resolution.

During our observations, the weather was very good in July~2017, with average system temperatures $T_{\rm sys}\simeq$120~K  and 150~K at 3~mm and 2~mm, respectively. It was more variable in May 2018, $T_{\rm sys}\simeq$(100-130)~K at 3~mm, and worse in August~2018,  $T_{\rm sys}\simeq$120~K and 200~K at 3~mm and 2~mm, respectively. The weather was good in January~2020, with $T_{\rm sys}\simeq$(105-120)~K, 200~K, and $\simeq(210-330)$~K at 3, 2, and 1~mm, respectively. We summarize our observations in Table 2. 
The data reduction and analysis were performed using the {\sc CLASS} software of the {\sc GILDAS}  package\footnote{https://www.iram.fr/IRAMFR/GILDAS/}. Our results are presented in Sect.~\ref{sec:results}.

\begin{table*}[htb]
\begin{center}
\begin{tabular}{cccll}
\hline\hline
Galaxy & CO(J$\rightarrow$J-1) &  integration time  &  Program ID      & Date                 \\
& & -- on-source, double polar -- & &  \\
 & & (hr) & &  \\
\hline
                &                      &                                            &                &         \\
A209   &  1$\rightarrow$0 & 0.9 & 173-17 &  May 2018  \\ 
       &   &  &  &    \\
M0329  &  1$\rightarrow$0 & 1.7 & 065-17 &  July 2017\\ 
       &  2$\rightarrow$1  & 1.7 & 065-17 & July 2017 \\ 
       &   &  &  &    \\
M0429  &  1$\rightarrow$0 & 0.4  & 065-17 &  July 2017  \\ 
       &  2$\rightarrow$1  & 0.4 & 065-17 & July 2017 \\
       &   &  &  &    \\
M0647  &  2$\rightarrow$1  & 3.5 & 088-18 & Aug 2018 \\
       &   &  &  &    \\
M0744  &   2$\rightarrow$1  & 1.2 & 065-17 & July 2017 \\ 
       &   &  &  &    \\
A611   &   1$\rightarrow$0   & 1.1 & 173-17  & May 2018 \\
       &   &  &  &    \\
M1115  &    1$\rightarrow$0  & 2.6 & 065-17 & July 2017 \\ 
       &   &  &  &    \\
A1423  &   1$\rightarrow$0  & 4.1  & 088-18 & Aug 2018 \\
       &   &  &  &    \\
M1206  &   1$\rightarrow$0  & 1.4 & 088-18 & Aug 2018 \\
       &   2$\rightarrow$1  & 1.4 & 088-18 & Aug 2018 \\
       &   &  &  &    \\
CL1226 &   4$\rightarrow$3  & 2.5 & 209-19 & Jan 2020 \\
       &   &  &  &    \\
M1311  &  1$\rightarrow$0  & 1.8  & 088-18 & Aug 2018 \\
       &  2$\rightarrow$1  & 1.8 & 088-18 & Aug 2018 \\
       &   &  &  &    \\
RX1347 &  1$\rightarrow$0  & 1.5 & 065-17 & July 2017 \\
       &  2$\rightarrow$1  & 1.5 & 065-17 & July 2017 \\
       &   &  &  &    \\       
M1423  &   1$\rightarrow$0  & 1.2 & 065-17 & July 2017 \\ 
       &   &  &  &    \\
RX1532 &   1$\rightarrow$0  & 2.0 & 065-17 & July 2017 and Jan 2020 \\
       &   3$\rightarrow$2  & 1.1 & 209-19 & Jan 2020 \\
       &   &  &  &    \\
M1720  &   1$\rightarrow$0  & 2.6  & 065-17 and 209-19 & July 2017 and Jan 2020\\ 
       &   2$\rightarrow$1  & 2.6 & 065-17 and 209-19 & July 2017 and Jan 2020\\ \\
       &   &  &  &    \\
A2261  &    1$\rightarrow$0  & 2.0 & 173-17 and 209-19 & May 2018 and Jan 2020\\ 
       &   &  &  &    \\
M2129  &   1$\rightarrow$0  & 3.3 & 088-18 & Aug 2018 \\ 
       &   2$\rightarrow$1  & 3.3 & 088-18 & Aug 2018 \\ 
       &   &  &  &    \\
RX2129 &   1$\rightarrow$0  & 1.7 & 173-17 & May 2018 \\ 
%
%
%

\hline
\end{tabular}
\caption{Summary of our IRAM 30m observations}
\label{tab:observinglog}
\end{center}
\end{table*}

\begin{table*}[tb]\centering
\begin{adjustwidth}{-0.7cm}{}
\begin{center}
\begin{tabular}{ccccccccccccc}
\hline\hline
 Galaxy  &  $z_{spec}$ & CO(J$\rightarrow$J-1)  & $\nu_{\rm obs}$ & $S_{\rm CO(J\rightarrow J-1)}\,\Delta\varv$   &  $M_{H_2}$ & $\tau_{\rm dep}$ & $\frac{M_{H_2}}{M_\star}$  & $\tau_{\rm dep, MS}$ & $\big(\frac{M_{H_2}}{M_\star}\big)_{\rm MS}$  \\
   &  & & (GHz) &  (Jy~km~s$^{-1}$)  & ($10^{10}~M_\odot$) & ($10^9$~yr) &  & ($10^9$~yr) & \\ 
 (1) & (2) & (3) & (4) & (5) & (6) & (7) & (8) & (9) & (10)  \\
 \hline
A209   &    0.206 & 1$\rightarrow$0 & 95.581 & $<2.2$ & $<2.0$  & --- & $<0.1$ & $1.29^{+0.21}_{-0.18}$ & $0.05^{+0.05}_{-0.05}$
\\
\hline
M0329  &    0.450 & 1$\rightarrow$0 & 79.497 & $<1.5$ & $<6.6$ & $<0.58$ & $0.092\pm0.028$ & $1.22^{+0.23}_{-0.20}$ & $0.08^{+0.08}_{-0.08}$ \\
       &          & 2$\rightarrow$1 & 158.992 & $2.4\pm0.7$ & $3.4\pm1.0$ & & &  & \\
\hline
M0429  &    0.399 & 1$\rightarrow$0 & 82.395  & $<2.5$ & $<8.7$ & $<1.1$ & $<0.07$ & $1.27^{+0.25}_{-0.21}$ & $0.07^{+0.06}_{-0.06}$ \\
       &          & 2$\rightarrow$1 & 164.788 & $<3.0$ & $<3.3$ & & &  & \\
\hline
M0647  &    0.591 & 2$\rightarrow$1 & 144.901 & $<1.7$ & $<4.5$ & --- & $<0.09$ & $1.19^{+0.25}_{-0.21}$ &  $0.10^{+0.09}_{-0.09}$\\
\hline
M0744  &    0.686 & 2$\rightarrow$1 & 136.737 & $<2.0$  & $<6.6$ & $<110$ & $<0.08$ & $1.19^{+0.28}_{-0.23}$  & $0.10^{+0.09}_{-0.09}$ \\
\hline
A611   &    0.288 & 1$\rightarrow$0 & 89.496 & $<1.0$  & $<1.8$ & --- & $<0.05$ & $1.30^{+0.24}_{-0.20}$ & $0.05^{+0.05}_{-0.05}$ \\
\hline
M1115  &    0.352 & 1$\rightarrow$0 & 85.260 & $<0.75$ & $<2.1$ & $<2.6$ & $<0.10$ & $1.21^{+0.20}_{-0.17}$ & $0.08^{+0.08}_{-0.08}$ \\
\hline
A1423  &    0.213 & 1$\rightarrow$0 & 95.030 & $1.7\pm0.5$ & $1.7\pm0.5$ & $<379$ & $0.087\pm0.026$ & $1.29^{+0.21}_{-0.18}$ & $0.05^{+0.05}_{-0.05}$ \\
\hline
M1206  &    0.440 & 1$\rightarrow$0 & 80.049 & $<1.2$ & $<5.4$ & $<12$ & $0.061\pm0.027$ & $1.26^{+0.26}_{-0.21}$ & $0.07^{+0.07}_{-0.07}$ \\
       &          & 2$\rightarrow$1 & 160.096 & $2.2\pm0.7$ & $3.0\pm1.0$ &  & &  & \\
\hline
CL1226 &    0.891 & 4$\rightarrow$3 & 243.808 & $<4.0$  & $<11.7$ & $<130$  & $<0.08$ & $1.17^{+0.32}_{-0.25}$ & $0.12^{+0.09}_{-0.09}$ \\
\hline
M1311  &    0.494 & 1$\rightarrow$0 &  77.156 & $<1.2$ & $<6.9$ & --- & $<0.05$ & $1.22^{+0.24}_{-0.20}$ & $0.09^{+0.08}_{-0.08}$ \\
       &          & 2$\rightarrow$1 & 154.309 & $<1.2$ & $<2.1$ & & &  & \\
\hline
RX1347 &    0.451 & 1$\rightarrow$0 & 79.443 & $<1.2$ & $<5.7$ & $<1.5$ & $<0.07$ & $1.22^{+0.23}_{-0.20}$ & $0.08^{+0.08}_{-0.08}$\\
       &          & 2$\rightarrow$1 & 158.882 & $<1.7$ & $<2.5$ & & &  & \\
\hline
M1423  &    0.545 & 1$\rightarrow$0 & 74.609 & $<0.75$ &  $<5.1$ & $<1.7$ & $<0.10$ & $1.20^{+0.25}_{-0.21}$ & $0.09^{+0.08}_{-0.08}$\\ 
\hline
RX1532 &    0.361 & 1$\rightarrow$0 & 84.696 & $3.0\pm0.4$ & $8.7\pm1.1$ &$0.73\pm0.16$ & $0.40\pm0.05$ & $1.21^{+0.20}_{-0.17}$ & $0.08^{+0.08}_{-0.08}$\\
       &          & 3$\rightarrow$2 & 254.075 & $20.2\pm1.6$ & & & &  & \\
\hline
M1720  &    0.391 & 1$\rightarrow$0 & 82.869 & $<2.2$ & $<7.5$ & $<11$  & $<0.08$ & $1.25^{+0.24}_{-0.20}$ & $0.07^{+0.07}_{-0.07}$\\
       &          & 2$\rightarrow$1 & 165.735 & $<2.9$  & $<3.0$ & & &  & \\
\hline
A2261  &    0.224 &  1$\rightarrow$0 & 94.176 &  $<1.8$  & $<2.0$  & --- & $<0.11$ & $1.27^{+0.20}_{-0.17}$ & $0.06^{+0.06}_{-0.06}$ \\
\hline
M1932  &    0.353 & 1$\rightarrow$0 & 85.260 & $3.5\pm0.5$ & $9.2\pm1.5$ & $0.45\pm0.09$ & $0.13\pm0.03$ & $1.35^{+0.29}_{-0.24}$ & $0.05^{+0.05}_{-0.05}$\\
       &          & 3$\rightarrow$2 & 255.766 & $29.0\pm2.8$ & & & &  & \\
       &          & 4$\rightarrow$3 & 341.007 & $33.7\pm3.5$ & & & &  & \\
\hline
M2129  &    0.570 & 1$\rightarrow$0 & 73.421 & $2.4\pm0.7$ & $17.9\pm5.2$ & $<1790$ & $0.73\pm0.22$ & $1.12^{+0.20}_{-0.17}$ & $0.13^{+0.11}_{-0.11}$\\
       &          & 2$\rightarrow$1 & 146.839 & $1.2\pm0.3$ & $5.6\pm1.4~\crosssymbol$ & & &  & \\
\hline
RX2129 &    0.234 & 1$\rightarrow$0 & 93.413 & $<1.0$ & $<1.2$ & --- & $<0.05$ & $1.29^{+0.21}_{-0.18}$ & $0.05^{+0.06}_{-0.05}$ \\
\hline
\end{tabular}
\end{center}
\caption{Molecular gas properties: (1) galaxy name;  (2) spectroscopic redshift as in Table~\ref{tab:BCG_properties}; (3-4) CO(J$\rightarrow$J-1) transition and observer frame frequency; (5) CO(J$\rightarrow$J-1) velocity integrated flux; (6) molecular gas mass obtained with $\alpha_{\rm CO}=4.36~M_\odot\,({\rm K~km~s}^{-1}~{\rm pc}^2)^{-1}$; (7) depletion timescale  $\tau_{\rm dep}=M_{H_2}/{\rm SFR}$; (8) molecular gas-to-stellar mass ratio; (9-10) depletion timescale and molecular gas-to-stellar mass ratio predicted for MS field galaxies with redshift and stellar mass of our targets, following \citet{Tacconi2018}. Upper limits are at 3$\sigma$. We refer to the text for further details. \\ \crosssymbol~The reported $M_{H_2}$ of M2129 is estimated from the CO(2$\rightarrow$1) flux and has been increased by a factor of two to take into account the possibility that the fit misses a substantial part of the CO(2$\rightarrow$1) emission.}
\label{tab:BCG_properties_mol_gas}
\end{adjustwidth}
\end{table*}

\section{Results}\label{sec:results}

\subsection{Molecular gas properties}\label{sec:IRAM30m_results}
The majority, that is, 72$\%, $ which is 13 out of 18 target BCGs, were not detected in CO by our observations. Four sources were tentatively detected in 
CO(1$\rightarrow$0) and/or CO(2$\rightarrow$1) with a signal-to-noise ratio (S/N) of $\simeq(3-4)$. Specifically, these were M0329, A1423, M1206, and M2129. We report their spectra in Fig.~\ref{fig:hint_detection_spectra}.

\begin{figure*}[]\centering
\captionsetup[subfigure]{labelformat=empty}
\subfloat[]{\hspace{0.cm}\includegraphics[trim={1cm 2cm 3.5cm 
5cm},clip,width=0.48\textwidth,clip=true]{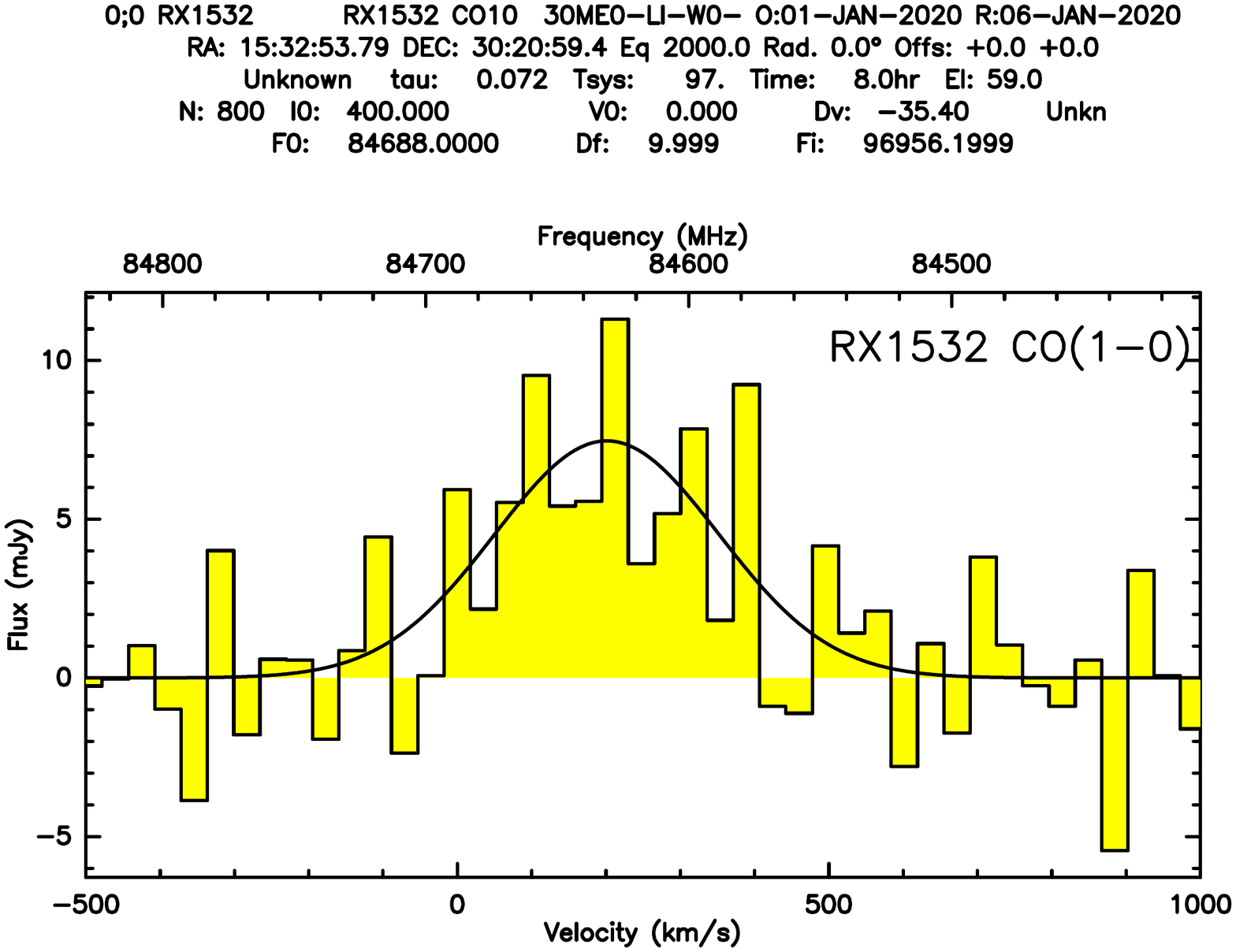}}
\subfloat[]{\hspace{0.7cm}\includegraphics[trim={1cm 2cm 3.5cm 
5cm},clip,width=0.48\textwidth,clip=true]{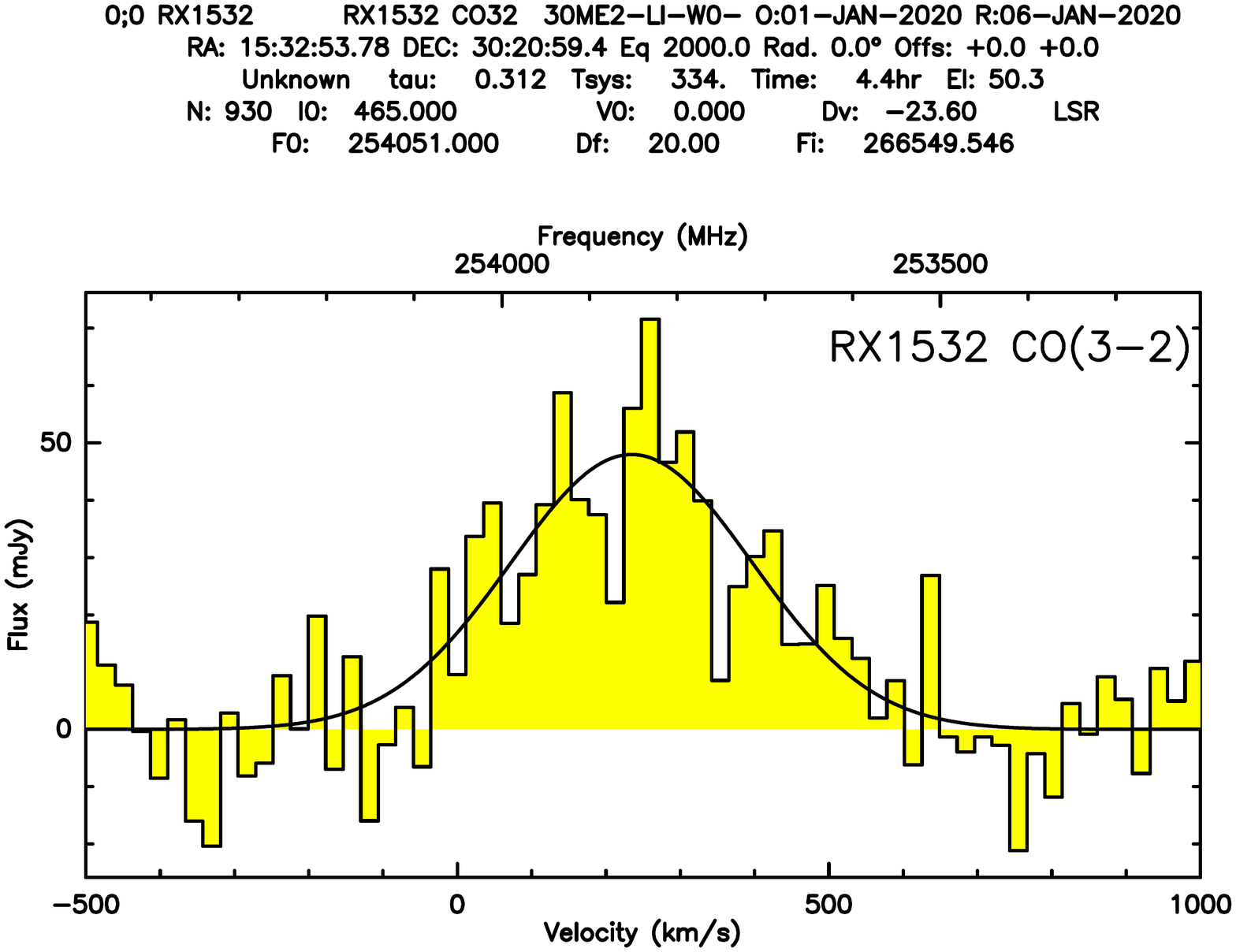}}\\
\caption{CO(1$\rightarrow$0) and CO(3$\rightarrow$2) baseline-subtracted spectra of the RX1532 BCG, obtained with the IRAM 30m. In both panels, the solid line shows the Gaussian best fit. In both panels the flux (y-axis) is plotted against the relative velocity with respect to the BCG redshift, as in Table~\ref{tab:BCG_properties} (bottom x-axis) and the observer frame frequency (top x-axis).}\label{fig:RX1532_spectra}
\end{figure*}

Our primary target,  RX1532, was unambiguously detected both in CO(1$\rightarrow$0) and CO(3$\rightarrow$2) at S/N=7.5 and 12.6, respectively, making it one of the few distant BCGs detected in CO, and possibly the only one in the distant Universe, that is, with $z>0.2$, along with M1932 \citep{Fogarty2019}, with clear detections in two different CO lines.  In Fig.~\ref{fig:RX1532_spectra}, we display the spectra of RX1532.

In Fig.~\ref{fig:HSTimages}, we present the {\it HST} images for the four BCGs with tentative detections in CO. The images show that these BCGs are massive elliptical galaxies with a bulge-dominated morphology, and are often surrounded by nearby companions that are likely hosted within the cluster cores. With an IRAM-30m beam of $\sim16$~arcsec at 2~mm \citep{Kramer2013}, we cannot exclude the possibility that nearby cluster core galaxies contribute to the total observed CO emission for these four BCGs. This is also suggested by the presence of offsets { of the order of a few $\sim$100~km/s}  that are seen in some CO spectra (Fig.~\ref{fig:hint_detection_spectra}). 

Interestingly, the {\it HST} image of M0647, that we observed in CO(2$\rightarrow$1) at $\sim$2~mm shows a nearby massive companion about $\sim5$~arcsec  toward the west from the BCG. The two sources are within the IRAM-30m beam. We therefore decided to point at the location of both galaxies, independently, and then average their resulting spectra. The CO results reported for M0647 in this work correspond to this  average. { We did not find any CO detection associated with the average, nor  with the two individual pointings.}

Based on the optical morphologies shown in Fig.~\ref{fig:RX1532_images} (left) for RX1532 and in Fig.~\ref{fig:HSTimages} for the four BCGs with hints of CO detections,
assuming a CO-to-optical size ratio $\sim0.5$ \citep{Young1995}, the target BCGs were  unresolved by our observations.
The fact that the sources are unresolved by our observations can be seen also in Fig.~\ref{fig:RX1532_images} (center and right), where the IRAM 30m beam at $\sim$1~mm is overplotted to the stellar mass and SFR maps of RX1532, also taking into account the fact that molecular gas is an excellent tracer of star formation   \citep{Bigiel2008,Schruba2011,Leroy2013}.

For the 13 BCGs with no detections in CO, we first removed the baseline from each spectrum by using a polynomial fit of degree one. Then we estimated rms noise levels for the antenna temperature (Ta$^\ast$).  For the five BCGs with CO detections, we instead fit each CO line with a Gaussian curve, after removing the baseline, as described above for the 13 BCGs with no detections in CO.

In Table~\ref{tab:BCG_properties_mol_gas}, we report the results of our analysis, where standard efficiency corrections were applied  to convert i) Ta$^\ast$ into the main beam temperature $T_{\rm mb}$ and then ii) $T_{\rm mb}$ into the corresponding CO line flux, where a 5~Jy/K conversion is used. { We adopted the following efficiency  corrections $T_{\rm mb}/T{\rm a}^\ast = 1.2$, $1.3$ and $1.6$ for $\sim$3, 2, and 1~mm observations, respectively.\footnote{https://www.iram.es/IRAMES/mainWiki/Iram30mEfficiencies}} In the case of non-detections, we estimated 3$\sigma$ upper limits to the integrated CO flux, at a resolution of 300~km/s, which corresponds to the typical width of CO lines of massive galaxies { and BCGs in particular \citep{Edge2001}.}

To derive the CO(J$\rightarrow$J-1) luminosity $L^{\prime}_{\rm CO(J\rightarrow J-1)}$, in units K~km~s$^{-1}$~pc$^2$, from the velocity integrated CO(J$\rightarrow$J-1) flux $S_{\rm CO(J\rightarrow J-1)}\,\Delta\varv\ $, in units Jy~km~s$^{-1}$, we used Eq.~(3) from \citet{Solomon_VandenBout2005}:
\begin{equation}
\label{eq:LpCO}
 L^{\prime}_{\rm CO(J\rightarrow J-1)}=3.25\times10^7\,S_{\rm CO(J\rightarrow J-1)}\,\Delta\varv\,\nu_{\rm obs}^{-2}\,D_L^2\,(1+z)^{-3}\,,
\end{equation}
where $\nu_{\rm obs}$ is the observer frame frequency, in GHz, of the CO(J$\rightarrow$J-1) transition, $D_L$ is the luminosity distance in Mpc, and $z$ is the redshift of the galaxy.

We then used the velocity integrated $L^{\prime}_{\rm CO(J\rightarrow J-1)}$ luminosities, or their upper limits, to derive estimates or $3\sigma$ upper limits to the total molecular gas mass $M_{H_2}=\alpha_{\rm CO}L^{\prime}_{\rm CO(1\rightarrow0)}=\alpha_{\rm CO}L^{\prime}_{\rm CO(J\rightarrow J-1)}/r_{J1}$. Here $r_{J1}= L^{\prime}_{\rm CO(J\rightarrow J-1)}/L^{\prime}_{\rm CO(1\rightarrow0)}$ is the excitation ratio. 
We assumed the following fiducial excitation ratios, typical of distant star-forming galaxies, namely, $r_{21}=0.8$  \citep[][]{Bothwell2013,Daddi2015,Freundlich2019}  and $r_{41}=0.4$ \citep[for CL1226,][]{Papadopoulos2000}. We note that RX1532 is the only source observed in CO(3$\rightarrow$2). However, since this BCG was detected also in CO(1$\rightarrow$0), we used the latter to estimate $M_{H_2}$. 

All the BCGs in our sample, except for RX1532 and M1932, have specific star-formation rates of ${\rm sSFR}\lesssim 3\times {\rm sSFR}_{\rm MS}$. For these 17 BCGs, we thus assumed a Galactic CO-to-H$_2$ conversion factor of $\alpha_{\rm CO}=4.36~M_\odot\,({\rm K~km~s}^{-1}~{\rm pc}^2)^{-1}$, which is typical of main sequence (MS) galaxies \citep{Solomon1997,Bolatto2013}. RX1532 and M1932 are more star-forming, with ${\rm sSFR}\gtrsim 5\times {\rm sSFR}_{\rm MS}$. To allow for a homogeneous comparison, we assumed the same Galactic CO-to-H$_2$ conversion factor also for these two sources. 
\citet{Fogarty2019} adopted instead a lower $\alpha_{\rm CO}=0.9~M_\odot\,({\rm K~km~s}^{-1}~{\rm pc}^2)^{-1}$ for M1932, which is, however, typical of ultra-luminous infrared galaxies \citep[e.g.,][for a review]{Bolatto2013}. { We  also note that  \citet{Geach2011} and \citet{Cybulski2016} adopted a Galactic CO-to-H$_2$ conversion for their moderate redshift galaxies ($z\sim0.2-0.4$), similarly to what has been done in this work. These sources are included in our comparison sample, and a fraction of them also have SFRs exceeding the MS, namely, ${\rm sSFR}\gtrsim 3\times {\rm sSFR}_{\rm MS}$, with estimated values up to $\sim10\,{\rm sSFR}_{\rm MS}$ \citep[see also Table~A.1 in][]{Castignani2020a}.}

\begin{table*}[]
\begin{center}
\begin{tabular}{cccccccccc}
\hline
Galaxy & CO(J$\rightarrow$J-1)  & $S_{\rm CO(J\rightarrow J-1)}\,\Delta\varv$ & S/N & FWHM & $L^\prime_{\rm CO(J\rightarrow J-1)}$ &  $z_{\rm CO(J\rightarrow J-1)}$ \\
 & & (Jy~km~s$^{-1}$) & & (km~s$^{-1}$) & ($10^{9}$~K~km~s$^{-1}$~pc$^2$) &   \\
 (1) & (2) & (3) & (4) & (5) & (6) & (7) \\

\hline
M0329 & 2$\rightarrow$1 & $2.4\pm0.7$ & 3.3 & $248\pm75$ & $6.3\pm1.8$ & $0.4508\pm0.0002$  \\ 
A1423 & 1$\rightarrow$0 & $1.7\pm0.5$ & 3.4  &  $886\pm306$ & $3.7\pm1.1$ & $ 0.2111\pm0.0006$ \\ 
M1206 & 2$\rightarrow$1 & $2.2\pm0.7$ & 3.1 & $266\pm72$ & $5.5\pm1.8$ & $0.4426\pm0.0002$\\ 
RX1532 & 1$\rightarrow$0 & $3.0\pm0.4$ & 7.5 & $369\pm51$ & $20.0\pm2.7$ & $0.3620\pm0.0001$\\ 
RX1532 & 3$\rightarrow$2 &  $20.2\pm1.6$ & 12.6 &   $390\pm36$ & $15.0\pm 1.2$& $0.3622\pm0.0001$ \\ 
M2129 & 1$\rightarrow$0 & $2.4\pm0.7$ & 3.4  & $312\pm121$ & $41\pm12$ & $0.5680\pm0.0002$ \\ 
M2129  & 2$\rightarrow$1 & $1.2\pm0.3$ & 4.0 &   $98\pm50$ & $10.2\pm2.6$~\cross & $0.5674\pm0.0001$\\
\hline
 \end{tabular}
\end{center}
 \caption{Summary of our IRAM 30m results for the sources with secure or tentative CO detections. Column description: (1) BCG name; (2) CO transition; (3) integrated CO line flux; (4) signal-to-noise ratio of the CO(J$\rightarrow$J-1) detection; (5) full width at half maximum of the CO(J$\rightarrow$J-1) line; (6) CO(J$\rightarrow$J-1) velocity integrated luminosity; (7) redshift derived from the CO(J$\rightarrow$J-1) line.\\ \crosssymbol~The reported $L^\prime_{\rm CO(2\rightarrow 1)}$ of M2129 is estimated from the CO(2$\rightarrow$1) flux and has been increased by a factor of two to take into account the possibility that the fit misses a substantial part of the CO(2$\rightarrow$1) emission.   } 
\label{tab:COresults}
\end{table*}



In Table~\ref{tab:COresults}, we summarize our IRAM 30m results for the five CLASH BCGs with CO detections and hints of detections  from our campaign. 
Due to the high S/N of the detected CO lines for RX1532 and the proximity in frequency of the CN(N=1$\rightarrow$0) doublet to the CO(1$\rightarrow$0) line, we used both LO and LI sidebands to look for CN(N=1$\rightarrow$0) J=1/2$\rightarrow$1/2 and CN(N=1$\rightarrow$0) J=3/2$\rightarrow$1/2 in RX1532, respectively, redshifted at $\sim83$~GHz in the observer frame. We did not find any of the two lines, which is not surprising, since they are usually much fainter than CO(1$\rightarrow$0).

We did not attempt to set any upper limit to the continuum emission of the target galaxies by using the available total 16~GHz (LI, LO, UI, UO) bandwidths,  for each polarization. In fact the faintness of our targets and the significant intrinsic atmospheric instability at mm-wavelengths prevented us from determining robustly the continuum level, or estimating its upper limit.

We used the SFRs and the molecular mass estimates, or their upper limits, to estimate depletion time scales $\tau_{\rm dep}=M_{H_2}/{\rm SFR}$, or their upper limits, associated with the consumption of the molecular gas.  Similarly, we  also estimated the molecular-gas-to-stellar-mass ratios, $M_{H_2}/M_\star$, or their upper limits, for the BCGs in our sample.  For comparison we computed the depletion time $\tau_{\rm dep, MS}$ and the molecular gas to stellar mass ratio $\big(\frac{M_{H_2}}{M_\star}\big)_{\rm MS}$ for MS field galaxies with redshift and stellar mass equal to those of our target galaxies, as found using empirical prescriptions by \citet{Tacconi2018}. The estimates are summarized in Table~\ref{tab:BCG_properties_mol_gas}.

Concerning M2129, in both Tables~\ref{tab:BCG_properties_mol_gas} and \ref{tab:COresults}, we increased by a factor of two both $L^{\prime}_{\rm CO(2\rightarrow 1)}$ and the associated H$_2$ mass.
We note however that for M2129, in this work we use the H$_2$ mass derived via $L^{\prime}_{\rm CO(1\rightarrow0)}$. 
The $L^{\prime}_{\rm CO(2\rightarrow 1)}$ correction  has been done  a posteriori, noting that the Gaussian fit reported in Fig.~\ref{fig:hint_detection_spectra} seems to miss a substantial part of the CO(2$\rightarrow$1) emission of M2129, which artificially results in a low excitation ratio $r_{21}$ for the galaxy. 
This correction yields an excitation ratio of $r_{21}\sim0.25$. For a comparison, we note that distant star-forming galaxies usually have a higher $r_{21}=0.84\pm0.13$ \citep{Bothwell2013}.

Concerning RX1532, which is the only other BCG that we detected in two different CO lines, we find that the gas is highly excited. An excitation ratio $r_{31}=0.75\pm0.12$ is found, which is higher than the value of $r_{31}\sim0.6$ typically observed in distant star-forming galaxies \citep{Devereux1994,Daddi2015} and similar to the value $r_{31}\sim0.92$ that \citet{Fogarty2019} found for M1932. 

Given the large number of upper limits to the molecular gas mass and the SFR of the BCGs in our sample we were able to set only upper limits, at $3\sigma$ level, to their depletion time. The only exceptions are RX1532 and M1932, which have robust estimates of both $M_{H_2}$ and the SFR, which allowed us to estimate their depletion time, as outlined in Table~\ref{tab:BCG_properties_mol_gas}. When deriving the upper limits reported in the Table, for each source we adopted the most stringent upper limit to $M_{H_2}$, in case of multiple values. Similarly, in the cases where both H$_\alpha$- and UV-based SFRs are available for a given BCG, we averaged between the two to estimate the depletion time, or its upper limit.

{ We also searched for CO emission by stacking the CO(1$\rightarrow$0) and CO(2$\rightarrow$1)  spectra, separately, of the BCGs with no individual detections. Among these BCGs, we excluded CL1226 for which we targeted CO(4$\rightarrow$3) only. We are left with 12 BCGs observed either in CO(1$\rightarrow$0) or in CO(2$\rightarrow$1), or both.
Our stacking yielded negative results, with 3$\sigma$ upper limits of $S_{\rm CO(1\rightarrow0)}\,\Delta\varv<$0.33~Jy~km~s$^{-1}$ and $S_{\rm CO(2\rightarrow1)}\,\Delta\varv<$0.64~Jy~km~s$^{-1}$, at a velocity resolution of 300~km/s. They imply upper limits of $M_{H_2}\lesssim1.0\times10^{10}~M_\odot$, estimated at the median redshifts of the two sets of CO(1$\rightarrow$0) and CO(2$\rightarrow$1) spectra that are used for the stacking. }



\begin{figure*}[h!]\centering
\subfloat{\hspace{0.2cm}\includegraphics[trim={0.cm 0cm 4.5cm 
0cm},clip,width=0.5\textwidth,clip=true]{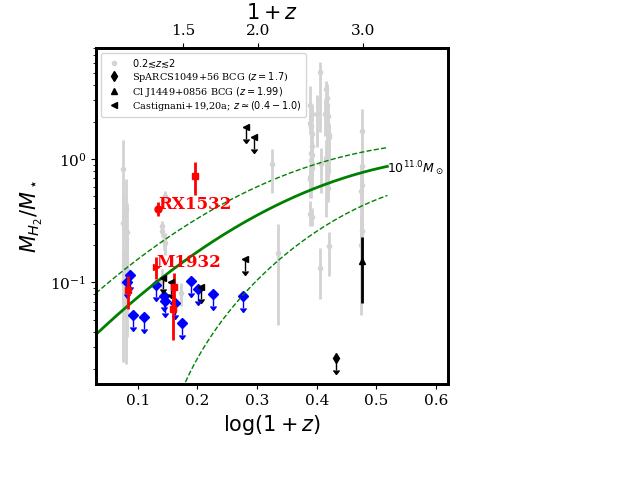}}
\subfloat{\hspace{0.2cm}\includegraphics[trim={0.cm 0cm 4.5cm 
0cm},clip,width=0.5\textwidth,clip=true]{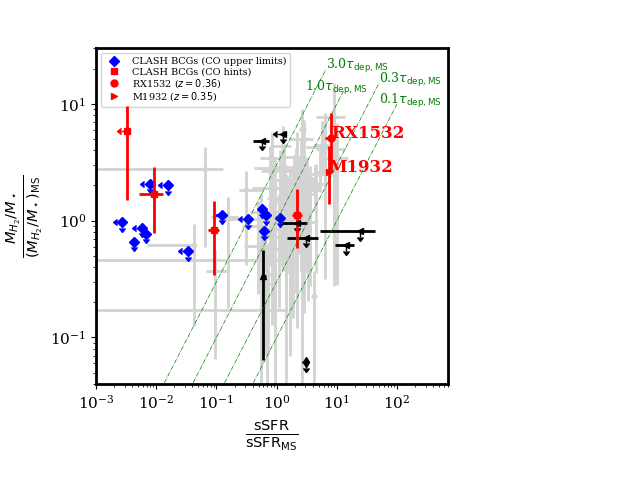}}\\
\caption{{ Molecular gas properties of distant  BCGs and cluster galaxies observed in CO}. Left: Evolution of the molecular gas-to-stellar mass ratio as a function of the redshift for cluster galaxies at $0.2\lesssim z\lesssim2$ observed in CO.  
The solid green curve is the scaling relation found  by \citet{Tacconi2018} for field galaxies in the MS and  with $\log(M_\star/M_\odot)$=11, which corresponds to the median stellar mass of all sources in the plot. The green dashed lines show the statistical 1$\sigma$ uncertainties in the model. Right: Molecular gas-to-stellar mass ratio vs. the specific SFR for the cluster galaxies in our sample, both normalized to the corresponding MS values using the relations for the ratio and the SFR by \citet{Tacconi2018} and \citet{Speagle2014}, respectively. The dot-dashed green lines show different depletion times, in units of the depletion time at the MS.}\label{fig:mol_gas1}
\subfloat{\hspace{0.2cm}\includegraphics[trim={0.cm 1.1cm 4.8cm 
1.2cm},clip,width=0.33\textwidth,clip=true]{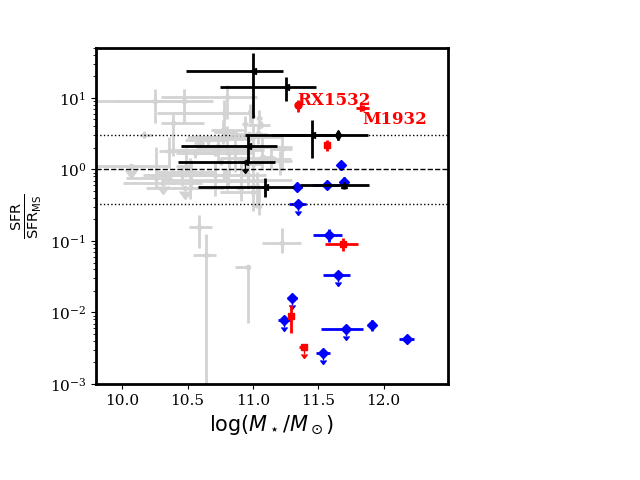}}
\subfloat{\hspace{0.2cm}\includegraphics[trim={0.cm 1.1cm 4.8cm 
1.2cm},clip,width=0.33\textwidth,clip=true]{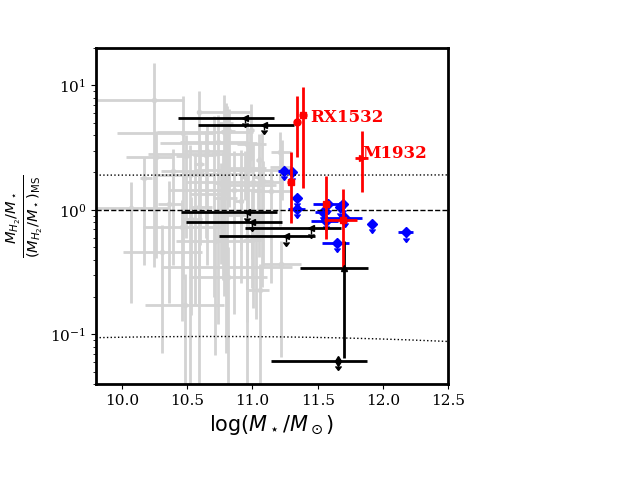}}
\subfloat{\hspace{0.2cm}\includegraphics[trim={0.cm 1.1cm 4.8cm 
1.2cm},clip,width=0.33\textwidth,clip=true]{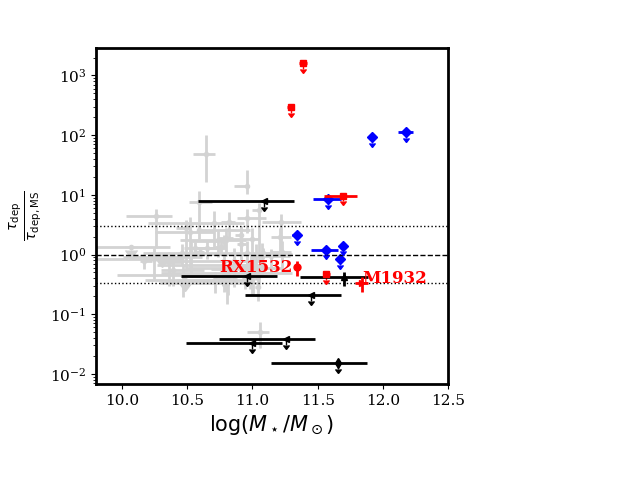}}
\caption{{ Star-formation rate (left), molecular gas-to-stellar mass ratio (center), and depletion time (right), as a function of the stellar mass for distant BCGs and cluster galaxies observed in CO.} The y-axis values are all normalized to the corresponding MS values using the relations by \citet{Speagle2014} and \citet{Tacconi2018}. { The horizontal dashed lines correspond to y-axis values equal to unity, while the dotted lines denote the fiducial uncertainties associated with the MS. { The uncertainty is chosen equal to $\pm0.48$~dex for both left and right panels, since the MS is commonly identified by $1/3<{\rm SFR}/{\rm SFR}_{\rm MS}<3$. For the central panel, the plotted uncertainties are estimated at redshift $z=0.5$.}} The color-coding for the data points is the same as in Fig.~\ref{fig:mol_gas1}.}\label{fig:mol_gas3}
\end{figure*}

\subsection{The comparison sample of distant cluster galaxies}\label{sec:comparison_sample}
We  compared the results found for our sources in terms of stellar mass, SFR, molecular gas content, and depletion time with those found in the literature for distant clusters galaxies with both stellar mass estimates and CO observations. 
For this comparison, we considered the compilation reported in \citet{Castignani2020a}, comprising 120 (proto-)cluster galaxies at $0.2\lesssim z\lesssim5.0$ with $M_\star>10^{9}~M_\odot$ and CO observations from the literature. Since our study is focused on massive distant BCGs of moderate to high ($z\lesssim1$) redshifts, we limit ourselves to sources in the compilation with stellar masses $M_\star>10^{10}~M_\odot$ and at $z<2$ (that is, we exclude the most distant clusters and still assembling proto-clusters). This selection yields 64 galaxies over 22 clusters at moderate to high redshift: $z\sim0.2-0.5$ \citep{Cybulski2016,Geach2011,Jablonka2013}; $z\sim0.4-1.0$ \citep{Castignani2019,Castignani2020a}; $z\sim1-1.2$ \citep{Wagg2012,Castignani2018}; $z\sim1.5-2.0$   \citep{Aravena2012,Rudnick2017,Webb2017,Noble2017,Noble2019,Hayashi2018,Kneissl2019,Coogan2018,Castignani2020b}.  We refer to \citet{Castignani2020a} and references therein for details.

Thanks to the additional inclusion of the 19 BCGs at $z\simeq0.2-0.9$ that are considered in this work, the full list of sources comprises 83 cluster galaxies over 41 clusters at $z\simeq0.2-2$
 and stellar masses in the range $\log(M_\star/M_\odot)\simeq10-12$.
Including ${\rm SFR}\gtrsim3\times{\rm SFR}_{\rm MS}$ sources might result in biased-high molecular gas masses \citep[see also e.g.,][for discussion]{Noble2017,Castignani2018,Castignani2019}.  However, since the two most abundantly star-forming BCGs in our CLASH sample, RX1532 and M1932, have ${\rm sSFR}\gtrsim 5\times {\rm sSFR}_{\rm MS}$, we prefer not to discard high-SFR galaxies from the comparison.

\subsection{Distant BCGs observed in CO}\label{sec:BCGs_CO_literature}
Among the distant 64 cluster galaxies with CO observations and stellar mass estimates considered in the previous section, there are only a few BCG candidates. As previously discussed in \citet{Castignani2019,Castignani2020a}, these are:
\begin{itemize}
 \item[$\bullet$] SpARCS~104922.6+564032.5 BCG at $z=1.7,$ which was originally observed in CO(2$\rightarrow$1) with the Large Millimiter Telescope (LMT). A large reservoir ($\sim10^{11}~M_\odot$) of molecular gas was detected, likely associated with a number of cluster core BCG companions, unresolved by the LMT, within the large 25~arcsec diameter (i.e., $\sim200$~kpc at $z = 1.7$). Our recent NOEMA observations \citep{Castignani2020b} confirmed this scenario, with the detection of two gas-rich BCG companions within 20~kpc from the BCG, and allowed us to set a robust upper limit to the molecular gas reservoir $M_{H_2}<1.1\times10^{10}~M_\odot$.\smallskip\smallskip\smallskip
\item[$\bullet$] ClJ1449+0856 BCG at $z = 1.99$ which is constituted by a gas-rich ($M_{H_2}\gtrsim10^{10}~M_\odot$) system (triplet) of galaxies.{ For this system, \citet{Coogan2018} detected both CO(4$\rightarrow$3) and CO(3$\rightarrow$2), possibly associated with an optically faint nearby Southern component \citep{Strazzullo2018}.} \smallskip\smallskip\smallskip
\item[$\bullet$] We\ also included for our comparison six additional BCG (candidates) at $z\sim0.4-1$ that were observed by \citet{Castignani2019,Castignani2020a} using the IRAM-30m as part of a large search for CO in distant star-forming BCGs, which were selected from several surveys (DES, SDSS, COSMOS, and SpARCS). These six sources are  DES-RG~399 ($z=0.39$), DES-RG~708 ($z=0.61$), COSMOS-FRI~16 ($z=0.97$), COSMOS-FRI~31 ($z=0.91$), 3C~244.1 ($z=0.43$), and  SDSS~J161112.65+550823.5 ($z=0.91$), for which we set robust upper limits to their molecular gas content in our previous studies. 
\end{itemize}


\begin{figure*}[h!]\centering
\subfloat{\hspace{0.2cm}\includegraphics[trim={0.cm 0.8cm 4.5cm 1.2cm},clip,width=0.45\textwidth,clip=true]{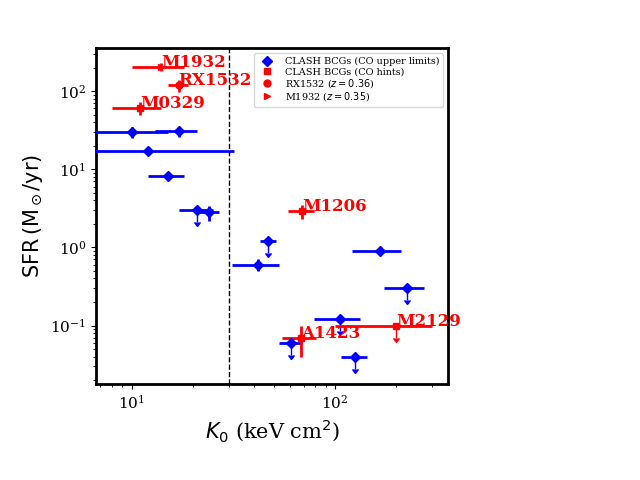}}
\subfloat{\hspace{0.2cm}\includegraphics[trim={0.cm 0.8cm 4.5cm 1.2cm},clip,width=0.45\textwidth,clip=true]{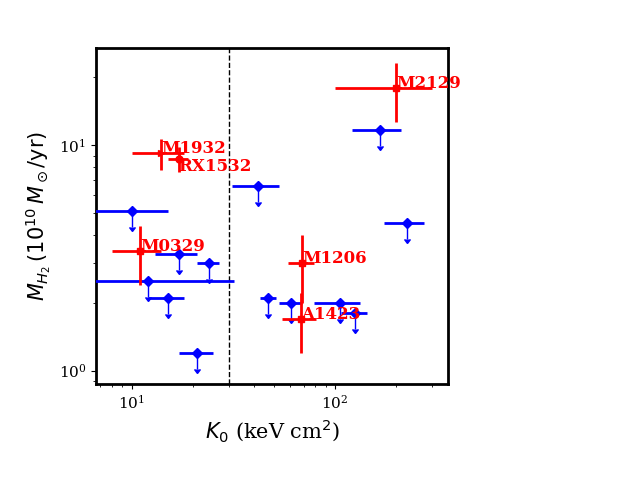}}
\caption{SFR (left) and molecular gas mass (right) vs. core entropy for CLASH BCGs with CO observations from our IRAM~30m campaign (this work), as well as for M1932, with ALMA observations by \citet{Fogarty2019}. The vertical dashed line is located at $K_0=30$~keV~cm$^2$. }\label{fig:entropy_plots}
\end{figure*}

In Fig.~\ref{fig:mol_gas1}, we show the ratio of molecular gas to stellar mass  $M_{H_2}/M_\star$ as a function of both redshift and specific SFR for the CLASH BCGs, as well as the comparison sample of distant sources with observations in CO and stellar mass estimates, as described above.  In Fig.~~\ref{fig:mol_gas3} we show instead the SFR, $M_{H_2}/M_\star$, and the depletion time ($\tau_{\rm dep})$, all normalized to their MS values, as a function of the stellar mass.
We distinguish in the figures the eight distant BCG (candidates), outlined above, from the remaining sources in the comparison sample.  The 19 CLASH BCGs considered in this work are also highlighted. We separately consider those with only upper limits (blue symbols) from those with detections in CO (red symbols).  We refer to the following section for  a discussion.

\section{Discussion}\label{sec:discussion}

\subsection{Stellar masses}
Figure~\ref{fig:mol_gas3} shows that the CLASH BCGs of our sample have exceptionally high stellar masses of $\log(M_\star/M_\odot)=11.6\pm0.2$, while the 56 normal sources with CO observations from the literature have a lower $\log(M_\star/M_\odot)=10.8^{+0.2}_{-0.3}$. Here, the median values and the 68\% confidence interval are reported.
We also compared the stellar masses of the 19 CLASH BCGs and all 27 BCGs with observations in CO, separately, with those of the 56 normal cluster galaxies in the comparison sample by means of the Kolmogorov-Smirnov test. The null hypothesis for the $\log(M_\star/M_\odot)$ distributions of the two populations is rejected with a significance of $7.1\sigma$ in both cases, p-values~$\simeq(1.1-1.7)\times10^{-12}$, which is consistent with the fact that BCGs are indeed the most massive galaxies in the clusters. 

\subsection{Molecular gas reservoirs}\label{sec:molgas}
Figure~\ref{fig:mol_gas1} shows that with our IRAM-30m observations we set robust upper limits of $M_{H_2}/M_\star\lesssim0.1$ for the 13 CLASH BCGs in our sample with no detections in CO, quite independently of the BCG redshift, within the entire $z\sim0.2-0.9$ range spanned by CLASH clusters.
Concerning the remaining six BCGs with detections in CO, RX1532 and M2129 are gas rich, with molecular gas to stellar mass ratios that are $\sim5-6$ times those of MS field galaxies. RX1532 and M2129 BCGs have indeed $M_{H_2}/M_\star\simeq0.4$, 0.7, respectively, corresponding to large gas reservoirs of $M_{H_2}\sim(1-2)\times10^{11}~M_\odot$. As displayed in Fig.~\ref{fig:mol_gas1}, they are thus among the distant cluster galaxies with the largest molecular gas reservoirs, also considering the fact that BCGs are the most massive galaxies in the clusters. {However, we stress here that some nearby galaxies are visible in the {\it HST} image of M2129 (Fig.~\ref{fig:HSTimages}), that fall within the IRAM-30m beam. Therefore, we cannot exclude the possibility that these companions contribute to the total observed CO emission.}

The other four BCGs (i.e., M0329, A1423, M1206, and M1932)
have instead  lower ratios of $M_{H_2}/M_\star\sim0.1$ and molecular gas reservoirs in the range $M_{H_2}\simeq(2-9)\times10^{10}~M_\odot$.  These values are similar to the upper limits in the range of $M_{H_2}\sim10^{10-11}~M_\odot$ that we set for the 13 CLASH BCGs with no detections in CO. These results suggest that these 13 BCGs tend to be gas poor, since we could have detected them with our IRAM-30m observations if they had cold gas reservoirs similar to those of the other CLASH BCGs with evidence of CO.

{ Among the four sources M1932 has a molecular gas content that is formally higher than that of MS field galaxies, that is $M_{H_2}/M_\star\simeq3\,(M_{H_2}/M_\star)_{\rm MS}$. However, as seen in Fig.~\ref{fig:mol_gas3} (center), it is still marginally consistent with that of the MS, once the uncertainties associated with the MS (horizontal dotted lines) and with the $M_{H_2}/M_\star$ ratio of M1932 are considered.}
The molecular-gas-to-stellar-mass ratios of { the remaining three} BCGs (M0329, A1423, and M1206) and the upper limits for the 13 BCGs are in agreement, within the uncertainties, with the values for MS field galaxies inferred using the empirical prescription by \citet{Tacconi2018}. 
Overall, the ratios, normalized to the MS, or their upper limits, are also in agreement with the values found for normal cluster galaxies with detections in CO, given the large scatter in the data points.

{ We also note that the median redshift of the 19 CLASH BCGs in our sample is $z_{\rm median}=0.4$, lower than the values of $z_{\rm median}=1.0$ and $1.5$ associated with the 22 clusters and 64 sources in the comparison sample, respectively. Similarly, the eight BCGs from previous studies that are considered for the comparison have $z_{\rm median}=0.9$. This suggests that in our study, we probe the molecular gas content of the still overlooked population of cluster core galaxies at moderate-to-high redshift. In fact, to some extent, we fill the existing gap at $z\sim0.2-0.9$, as appreciated from Fig.~\ref{fig:mol_gas1} (left). Furthermore, while there is an overall increase of the $M_{H_2}/M_\star$ ratio with redshift out to $z\sim2$ (Fig.~\ref{fig:mol_gas1}, left), once the ratio is normalized to the MS value, for the comparison sample we do not see any statistically significant bias towards higher $\frac{M_{H_2}/M_\star}{(M_{H_2}/M_\star)_{\rm MS}}$ values  than those of the CLASH BCGs (Fig.~\ref{fig:mol_gas1}, right).}

Although the molecular gas content of the CLASH BCGs considered in this work is generally consistent with that of MS field galaxies and distant cluster galaxies observed in CO, we stress here that the upper limits of $M_{H_2}/M_\star\lesssim0.1$ we set for the subsample of 13 CLASH BCGs are among the lowest ratios found for distant cluster galaxies, and BCGs in particular. This can be appreciated in Fig.~\ref{fig:mol_gas1}. For a comparison, BCGs candidates from our recent IRAM-30m campaign \citep{Castignani2019,Castignani2020a}, at similar redshifts of $z\sim0.4-0.1$ than CLASH BCGs,  indeed have upper limits ranging from $M_{H_2}/M_\star\lesssim2$ (COSMOS-FRI~16, 31 at $z\sim0.9-1$), down to  lower values of $M_{H_2}/M_\star\sim0.1-0.2$ (SDSS~J161112.65+550823.5, 3C~244.1, DES-RG~399, and DES-RG~708 at $z\sim0.4-0.9$). SpARCS~104922.6+564032.5 BCG, at $z=1.7,$ which we included in our comparison sample represents a remarkable exception. Thanks to recent interferometric  NOEMA observations we inferred a low upper limit of $M_{H_2}/M_\star<0.02$ to the molecular gas to stellar mass ratio, which is among the lowest ones observed for distant $z > 1$ ellipticals \citep[e.g.,][]{Sargent2015,Gobat2018,Bezanson2019}.


\subsection{Star formation and depletion time}\label{sec:SFR_tdep}
The CLASH BCGs in our sample span a broad range of star formation from low values of SFR$\lesssim1~M_\odot$/yr up to SFR$\gtrsim100~M_\odot/$yr.
Interestingly, the three most star-forming BCGs, M1932, RX1532, and M0329, with ${\rm SFR}>{\rm SFR}_{\rm MS}$ and both H$_\alpha$ and UV based SFR$\gtrsim40~M_\odot/$yr, have all been detected in CO at S/N$>3$.

Concerning the remaining sources, three out of four BCGs with tentative detections in CO (i.e., A1423, M1206, M2129) and the 13 BCGs with only upper limits in CO exhibit star formation activity ranging from MS values ${\rm SFR}\sim{\rm SFR}_{\rm MS}$ down to a few $10^{-3}{\rm SFR}_{\rm MS}$. These results suggest that tentative detections in CO as well as non-detections are found quite independently of the star formation activity of the BCG. 

However, A1423, M1206, and M2129 with tentative detections in CO have low star-formation activity SFR$\sim(1-3)~M_\odot$/yr, corresponding to low normalized values of $\sim(0.009-0.09){\rm SFR}_{\rm MS}$, which are rarely seen associated with cluster galaxies detected in CO (see Fig.~\ref{fig:mol_gas1}, right). The presence of flows of cold gas could possibly explain the existence of large molecular gas reservoirs in these BCGs. However, the three sources are hosted in dynamically disturbed clusters with high level of central entropy $K_0\sim(70-200)$~keV~cm$^2$ \citep{Donahue2015,Donahue2016}, which make this scenario less plausible. 
Follow-ups in CO are needed to reject or confirm with higher significance the tentative detections for these BCGs.

We used the SFRs and the molecular mass estimates, or their upper limits, to estimate the depletion time scales $\tau_{\rm dep}=M_{H_2}/$SFR for the 19 CLASH BCGs. We set upper limits to the depletion time for all 13 BCGs with non-detections and for the 4 BCGs with only tentative detections in CO, due to the large uncertainties associated with the SFR and the relatively low S/N associated with their CO detections. Upper limits to $\tau_{\rm dep}$ span a broad range, from relatively low values $\lesssim1-2$~Gyr (e.g., M0329, M0429, M115, RX1347, and M1423) to very high values, in some cases even formally exceeding the Hubble time (e.g., M0744, A1423, CL1226, M2129). We refer to Table~\ref{tab:BCG_properties_mol_gas} for details.

The two CLASH BCGs with secure CO detections, RX1532 and M1932, have relatively low depletion time scales of $\tau_{\rm dep}\simeq$~0.7 and 0.5~Gyr, respectively. Although the time scales are formally consistent with those of MS galaxies, as seen in Figs.~\ref{fig:mol_gas1}, \ref{fig:mol_gas3}, the two BCGs tend to populate the regions associated with low depletion times and high SFRs that are occupied by other rare distant star-forming BCGs \citep[e.g.,][]{Castignani2019,Castignani2020a}. 


\subsection{Star formation and AGN activity}\label{sec:SFR_AGN}
AGN activity might result in biased-high SFRs and biased-low depletion time scales \citep[e.g.,][for further discussion]{Castignani2020a}. The BCGs in the comparison sample are indeed almost invariably associated with radio galaxies with rest frame 1.4~GHz luminosity densities in the range $L_{\rm 1.4~GHz}\simeq10^{31-34}$~erg~s$^{-1}$~Hz$^{-1}$ \citep{Castignani2019,Castignani2020a,Trudeau2019}. 
Similarly, \citet{Yu2018} studied the low-frequency radio properties of a sample of 20 relaxed CLASH clusters using JVLA observations at 1.5~GHz. The authors did not find any clear trend between the cluster central entropy $K_0$ and the radio power. However, in the case of cool-core clusters, that is, $K_0<30$~keV~cm$^2$, the authors found radio luminosities mostly in the range $\sim10^{31-32}$~erg~s$^{-1}$~Hz$^{-1}$. This also applies to the nine CLASH BCGs in our sample associated with $K_0<30$~keV~cm$^2$ cluster cores. Among the nine sources, there are the three most star-forming BCGs in our sample: M0329, with a tentative CO detection, as well as RX1532 and M1932, with secure detections in CO.  These have rest frame 1.5~GHz luminosity densities of $(2.37\pm0.42)\times10^{31}$~erg~s$^{-1}$~Hz$^{-1}$, $(6.44\pm0.08)\times10^{31}$~erg~s$^{-1}$~Hz$^{-1}$, and $(7.65\pm0.02)\times10^{31}$~erg~s$^{-1}$~Hz$^{-1}$, respectively \citep{Yu2018}. Similarly, the remaining six BCGs in our sample with $K_0<30$~keV~cm$^2$ with only upper limits in CO (namely, M0429, M1115,  RX1347, M1423,  M1720, and RX2129) have luminosity densities between $3\times10^{31}$~erg~s$^{-1}$~Hz$^{-1}$ and $6\times10^{32}$~erg~s$^{-1}$~Hz$^{-1}$.

However, we stress that all CLASH BCGs in our sample have accurate SFR estimates based on UV analysis \citep{Fogarty2015,Donahue2015}, as summarized in Table~\ref{tab:BCG_properties} and illustrated in Fig.~\ref{fig:RX1532_images} for RX1532. All nine BCGs with $K_0<30$~keV~cm$^2$ also have H$_\alpha$-based SFRs or upper limits, as in the case of RX2129. Therefore, despite the fact that CLASH BCGs are almost invariably associated with radio-loud AGNs, these aspects strengthen the reliability of the SFR estimates used in this work.

\subsection{Central entropy, star formation, and molecular gas}
The BCGs in our sample are hosted in clusters with different levels of central entropy $K_0$, estimated from X-ray analysis by \citet{Donahue2015}, see Table~\ref{tab:BCG_properties}. As discussed in the previous section, CLASH BCGs found in low-entropy cluster cores are associated with radio galaxies with rest frame 1.5~GHz luminosity densities of $\sim10^{31-32}$~erg~s$^{-1}$~Hz$^{-1}$. This suggests that the cooling of the ICM may result in the infall of gas that ultimately accretes onto the central supermassive black holes and sustain the production of radio emitting jets. However, the infalling gas may also replenish the BCGs' reservoirs and ultimately sustain high levels of star formation. The recent study by \citet{Fogarty2017} indeed supports this scenario. The authors show, in fact, a remarkable tight anti-correlation, with a limited scatter of 0.15~dex, between the SFR and $t_{\rm cool}/t_{\rm ff}$ for the sample of CLASH BCGs. Here $t_{\rm cool}/t_{\rm ff}$ is the ICM cooling time to free-fall time ratio, which can be used as proxy for the thermal instability of ICM gas. If the gas cools quickly with respect to the time it takes to infall towards the cluster center, it can become thermally unstable and then collapse, as suggested by simulations \citep[e.g.,][]{Gaspari2012,Li_Bryan2014,Li2015}.
Observations have also confirmed this scenario. In fact, multiphase infalling gas, both atomic and molecular, is often found within the central low-entropy and short-cooling-time regions around local BCGs, with $t_{\rm cool}\lesssim20\times t_{\rm ff}$ \citep{Olivares2019,Voit_Donahue2015,Voit2017}.

Inspired by these studies, similarly to \citet{Fogarty2017}, in Fig.~\ref{fig:entropy_plots} (left), we plot the SFR vs. $K_0$ for the 19 CLASH BCGs in our sample.  Since the molecular gas content is correlated with the ongoing star formation activity \citep{Bigiel2008,Schruba2011,Leroy2013}, in the right panel of the figure we show also the molecular gas mass $M_{H_2}$ versus the central entropy $K_0$ for the 19 sources. 
Analogously to \citet{Fogarty2017}, the plot in the left panel shows that SFR and the central entropy are anti-correlated, with a significance of 4.2$\sigma$ (p-value~$=2.5\times10^{-5}$), as found with the Spearman test, where upper limits were considered as true measurements. { A lower significance of 2.5$\sigma$ (p-value~$=0.012$) is found in the case where SFR upper limits are discarded.}

{{ No clear trend is instead} observed between the molecular gas content { (expressed either as  $M_{H_2}$ or as $M_{H_2}/M_\star$)}  and $K_0$, when considering either the full sample of 19 BCGs or the subsample of six BCGs with tentative or secure detections in CO (Fig.~\ref{fig:entropy_plots}, right). The 13 BCGs that are not detected in CO span quite uniformly the range $K_0\simeq10-200$~KeV~cm$^2$ (Fig.~\ref{fig:entropy_plots}); they have upper limits between $M_{H_2}\lesssim10^{10}~M_\odot$ and $M_{H_2}\lesssim10^{11}~M_\odot$, corresponding to  $M_{H_2}/M_\star\lesssim0.1$, quite independently of the central entropy.}

{ However,} BCGs with low entropy levels tend to have larger molecular gas reservoirs. In fact, limiting ourselves to the subsample of six BCGs with tentative or secure detections in CO, the three sources RX1532, M1932, and M0329, which are hosted in $K_0<30$~KeV~cm$^2$ cluster cores, seem to have a higher molecular gas content $M_{H_2}\simeq(3-10)\times10^{10}~M_\odot$ than M1206 and A1423, with $M_{H_2}\simeq(2-3)\times10^{10}~M_\odot$, that have high central entropy levels $K_0>30$~KeV~cm$^2$. M2129 is an exception, with a large molecular gas reservoir $M_{H_2}\sim2\times10^{11}~M_\odot$ and also a large central entropy level $K_0\simeq200$~KeV~cm$^2$. However, as discussed in Sect.~\ref{sec:molgas} nearby companions may contribute to the total CO emission for M2129.




These results suggest that the interplay between i) the star formation activity; ii) ICM cooling; iii) molecular gas reservoirs feeding the star formation; and iv) the stellar content of the BCGs is more complex than that observed between the SFR and the ICM cooling. We refer to the following section for further discussion.

\subsection{The case of RX1532: a rare gas-rich star-forming BCG}\label{sec:RX1532}

The three most abundantly star-forming BCGs in our sample are not only detected in CO at S/N$>3,$ but they also unambiguously exhibit low central entropy values of $K_0<30$~KeV~cm$^2$ (see also Sects.~\ref{sec:SFR_tdep}, \ref{sec:SFR_AGN}). Furthermore, the only two BCGs with clear detections in CO, M1932 and RX1532, are those with the highest SFR$\gtrsim100~M_\odot$/yr. These results suggest that the detection of large reservoirs of molecular gas in distant BCGs is possible when the following two conditions are met: i) high levels of star formation activity; and ii) cool-core cluster environments, which favor the condensation and the inflow of gas onto the BCGs themselves. A possible indirect evidence of such a gas condensation is the detection for both M1932 and RX1532 of highly excited gas $r_{31}\sim0.8-0.9$ for the high J=3$\rightarrow$2 CO transition, which may require the need of high gas densities $n_{H_2}\sim10^{4-5}~{\rm cm}^{-3}$ \citep[e.g.,][]{Riechers2011,Papadopoulos2012}. 


Consistently with this scenario, although our single-dish observations do not allow us to spatially resolve the observed CO emission for RX1532,  filamentary structures around the BCG are detected by {\it HST} in UV, H$_\alpha$, and in infrared with the WFC3-IR filter (Fig.~\ref{fig:RX1532_images}), while clumpy substructures around the BCGs have been found with our morphological {\sc Galfit} analysis (Fig.~\ref{fig:RX1532_galfit}). Higher resolution observations in CO could allow us to spatially resolve the BCG and have better insights on the morphology of its molecular gas reservoir.
Similarly,  as further discussed in Sect.~\ref{sec:introduction}, using ALMA observations at a resolution $<1''$, \citet{Fogarty2019} detected  CO(1$\rightarrow$0), as well as higher-J CO(3$\rightarrow$2) and CO(4$\rightarrow$3) transitions associated with a compact component associated with the host galaxy as well as a more diffuse component cospatial with UV knots and H$_\alpha$ filaments observed towards northeast by {\it HST}.

As observed in Fig.~\ref{fig:entropy_plots}, among the three BCGs (M1932, RX1532, M0329) with low values of $K_0<30$~KeV~cm$^2$ and high SFR$\gtrsim40~M_\odot$/yr, which are also detected in CO, it is RX1532 that seems to be the most gas rich, with $M_{H_2}/M_\star=0.40\pm0.05$. For a comparison, we note that RX1532 and M1932 have a similar molecular gas content $L^\prime_{\rm CO(1\rightarrow0)}\simeq2\times10^9$~K~km~s$^{-1}$~pc$^2$ and a similar global star-formation activity (SFR$\gtrsim100~M_\odot$/yr). However the molecular gas to stellar mass ratio of M1932 ($M_{H_2}/M_\star=0.13\pm0.03$) is a factor of $\sim3$ lower than that of RX1532.

The main reason for such a difference is that RX1532 has a lower stellar mass than M1932 by a factor of $\sim3$. It is worth mentioning that the stellar masses of all CLASH BCGs in our sample, except for M1932, have been taken from \citet{Burke2015}, who corrected the estimates for the contribution by the intra-cluster light (ICL). While specifically considering the sample of CLASH BCGs, \citet{Burke2015} did not report any stellar mass estimate for M1932 and we therefore used for this source the estimate by \citet{Cooke2016}. Since \citet{Cooke2016} did not consider the contribution of the ICL it is possible that the stellar mass of M1932 has a systematic offset with respect to those of other BCGs in our sample. However, as a sanity check, \citet{Cooke2016} performed a comparison between their stellar mass estimates and those by \cite{Burke2015} for the CLASH BCGs. They found a good agreement between the two, within 30\%. This result suggests that the discrepancy between the $M_{H_2}/M_\star$ ratio of RX1532 and M1932 is not due to the different methods used to estimate their stellar mass. In fact, if we adopt for RX1532 the stellar mass of $M_\star=(2.88\pm0.06)\times10^{11}~M_\odot$ by \citet{Cooke2016} we still find a high value of $M_{H_2}/M_\star=0.30\pm0.04$, only $\sim25\%$ smaller than that estimated using the RX1532 stellar mass by \citet{Burke2015}.

Our results suggest, thus, that RX1532 belongs to the rare population of star-forming and gas-rich BCGs in the distant universe. Furthermore, our systematic search for CO among the CLASH sample also shows that distant BCGs are generally gas-poor,  $M_{H_2}/M_\star\lesssim0.1$,  similarly to what has been found for other distant BCGs \citep[e.g,][]{Castignani2019,Castignani2020a,Castignani2020b,Coogan2018} and quite independently of the redshift considered, within the range of $z\sim0.2-0.9$ considered.  With this work, we thus provide new insights about the molecular gas content of the still overlooked population of cluster core galaxies of moderate to high-redshifts, in the range of $z\sim0.2-0.9$ ($\sim5$~Gyr of cosmic time). In this redshift range, there are in fact only a few observations of CO for cluster galaxies  \citep{Cybulski2016,Geach2011,Jablonka2013,Castignani2020c} and BCG (candidates) in particular \citep{Castignani2019,Castignani2020a}. This is nevertheless an important epoch in the evolution of galaxies, where the cosmic SFR density increases by a factor of $\sim3$ \citep{Madau_Dickinson2014} and, at least for field galaxies, the molecular gas to stellar mass ratio $M_{H_2}/M_\star\sim0.1\,(1+z) ^2$ is expected to increase by a similar factor of $\sim2.5$ \citep[][]{Carilli_Walter2013}. 


\section{Conclusions}\label{sec:conclusions}
In this work, we investigate the effect of the cluster environment in processing the molecular gas content in distant brightest cluster galaxies (BCGs), as part of a large search for CO in distant BCGs \citep{Castignani2019,Castignani2020a,Castignani2020b}. 
To this aim, we used the IRAM-30m telescope to observe in CO a sample of 18 BCGs at $z\sim0.2-0.9$, mostly in the northern hemisphere, drawn from the sample of 25 clusters in the  Cluster Lensing And Supernova survey with Hubble (CLASH) survey \citep{Postman2012a}. 
CLASH clusters have high-resolution, multi-band {\it HST} observations, deep X-ray monitoring with {\it Chandra}, and have been targeted with several spectroscopic campaigns, which make them excellent laboratories to investigate the evolution of galaxies in dense Mpc-scale environments, and BCGs in particular, as well as their interplay with the intra-cluster medium.

The sources were observed in several CO(J$\rightarrow$J-1) lines, with J~=~1, 2, 3, or 4. For each source, the specific CO lines were chosen to maximize the likelihood for a detection after taking the BCG redshift into account. In the sample, we also included the BCG M1932, in the southern hemisphere, which belongs to the CLASH sample and which  was recently detected with ALMA in several CO lines by \citet{Fogarty2019}.

An analysis of the {\it HST} I-band image and light profile of RX1532 carried out with {\sc Galfit} \citep{Peng2002,Peng2010} yields a half-light radius of $r_e=(11.6\pm0.3)$~kpc, which is formally higher than that predicted for field galaxies of similar mass and redshift. Our analysis also reveals a complex morphology for the BCG, with several clumpy substructures and a filamentary network extending from the host galaxy out to $\sim${ 20}~kpc, well outside the half-light radius. These results are consistent with the presence of UV knots and H$_\alpha$ filaments previously observed for RX1532 by {\it HST} \citep{Fogarty2015}.

With our IRAM-30m observations, we also unambiguously detected RX1532 both in CO(1$\rightarrow$0) and CO(3$\rightarrow$2) at S/N=7.5 and 12.6, 
respectively. By assuming a Galactic CO-to-H$_2$ conversion, we found that the source hosts a large reservoir of cold gas $M_{H_2}=(8.7\pm1.1)\times10^{10}~M_\odot$. Our observations also imply that the gas is highly excited, with $r_{31}=0.75\pm0.12$, which is higher than the value of $r_{31}\sim0.6$ typically found for distant star-forming galaxies \citep{Devereux1994,Daddi2015}.
Higher resolution observations in CO could allow us to spatially resolve the molecular gas content of RX1532. 
By looking for compact and more elongated components (e.g., tails, filaments, companions) we could distinguish between different environmental processing mechanisms (e.g., mergers, flows of cold gas, ram pressure stripping, and tidal stripping).

Among the 18 BCGs we observed with the IRAM-30m, it is only RX1532 that we securely detected in CO. However, we tentatively detected CO lines at S/N$\sim3-4$ in four additional BCGs at $z\sim0.2-0.6$. They include M0329, M1206, and A1423, with  $M_{H_2}\simeq(2-3)\times10^{10}~M_\odot$ and $M_{H_2}/M_\star\sim0.1$. We also tentatively detected M2129, both in CO(1$\rightarrow$0) and CO(2$\rightarrow$1), which tends to be more gas-rich, with $M_{H_2}\simeq2\times10^{11}~M_\odot$ and $M_{H_2}/M_\star\sim0.7$. The four BCGs also cover a broad range of star formation activity, ranging from $\sim(0.1-3)~M_\odot$/yr  (M1206, M2129, and A1423) to  high levels of star formation $\gtrsim40~M_\odot$/yr (M0329). 
For the remaining 13 BCGs, we set robust upper limits to the molecular gas to stellar mass ratio $M_{H_2}/M_\star\lesssim0.1$, which are among the lowest ratios found for distant ellipticals \citep[e.g.,][]{Sargent2015,Gobat2018,Bezanson2019} and BCGs in particular \citep{Castignani2019,Castignani2020a,Castignani2020b}.

By comparison with a sample of distant  cluster galaxies observed in CO and with the empirical model by \citet{Tacconi2018} for field galaxies, our results also suggest that RX1532 ($M_{H_2}/M_\star=0.40\pm0.05$) belongs to the rare population of star-forming and gas-rich BCGs in the distant universe.
Finally, we compared the molecular gas content of the BCGs with both their star formation activity and the central entropy of the intra-cluster medium, as inferred from X-ray analysis \citep{Donahue2015}. The comparison suggests that the three most star-forming BCGs in our sample (RX1532, M1932, and M0329, with SFR$>40~M_\odot$/yr) are not only detected in CO at S/N$>3$ but  also unambiguously exhibit low central entropy values of $K_0<30$~KeV~cm$^2$.
These results imply that the detection of large reservoirs of molecular gas $\gtrsim10^{10}~M_\odot$ in distant BCGs is possible when the following two conditions are met: i) high levels of star formation activity and ii) cool-core cluster environments, which favor the condensation and the inflow of gas onto the BCGs themselves. This is similar to what has previously been found for some local BCGs \citep[e.g.,][]{Salome2006,Tremblay2016}.

\begin{acknowledgements}
{ We thank the anonymous referee for helpful comments which contributed to improving the paper.}
This work is based on observations carried out under project numbers 065-17, 173-17, 088-18, and 209-19 with the IRAM 30m telescope. GC, MP, and SH thank IRAM staff at Granada for their help with these observations. IRAM is supported by INSU/CNRS (France), MPG (Germany) and IGN (Spain). This publication has made use of data products from the NASA/IPAC Extragalactic Database (NED).
GC acknowledges financial support from both the Swiss National Science Foundation (SNSF) and the Swiss Society for Astrophysics and Astronomy (SSAA). GC acknowledges the hospitality of the Observatory of Paris in Feb 2020, that helped to finalize the present work. MP thanks CEFIPRA organization for funding support (project no. 5402-2). 
\end{acknowledgements}

\begin{figure*}[h!]\centering
\captionsetup[subfigure]{labelformat=empty}
\subfloat[]{\hspace{0.cm}\includegraphics[trim={1cm 0cm 0cm 0cm},clip,width=0.45\textwidth,clip=true]{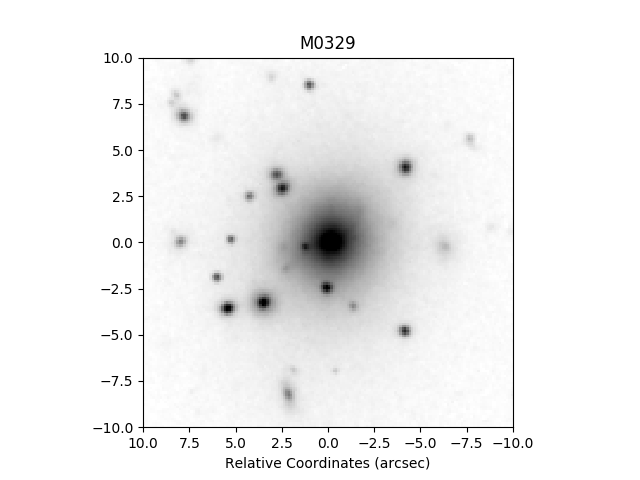}} 
\subfloat[]{\hspace{0.cm}\includegraphics[trim={1cm 0cm 0cm 0cm},clip,width=0.45\textwidth,clip=true]{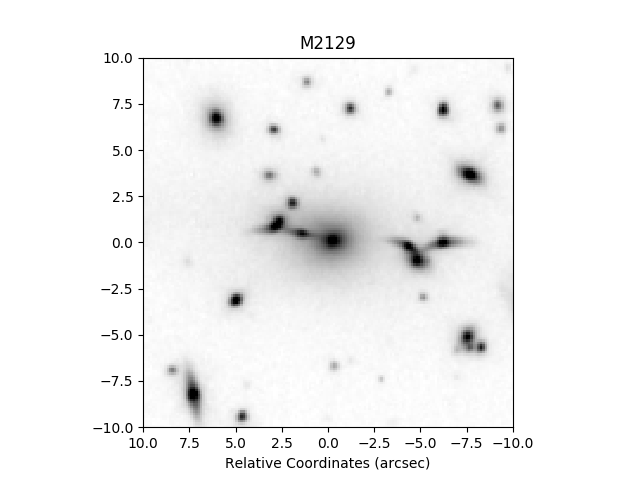}}\\ 
\subfloat[]{\hspace{0.cm}\includegraphics[trim={1cm 0cm 0cm 0cm},clip,width=0.45\textwidth,clip=true]{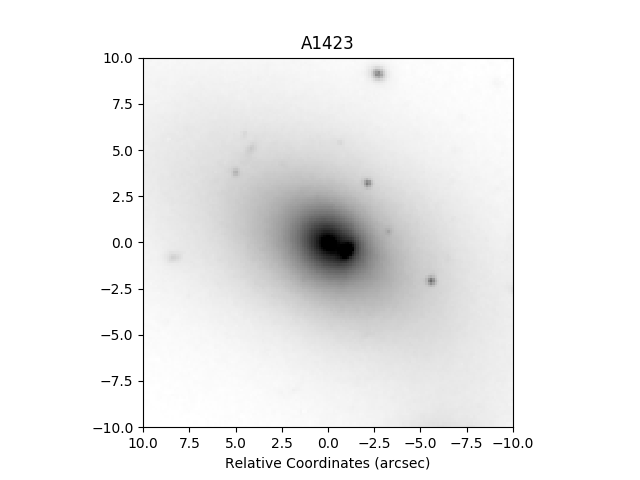}} 
\subfloat[]{\hspace{0.cm}\includegraphics[trim={1cm 0cm 0cm 0cm},clip,width=0.45\textwidth,clip=true]{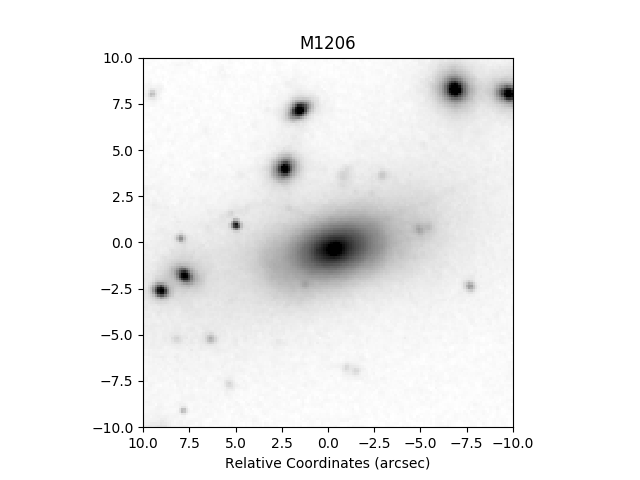}}\\ 
\subfloat[]{\hspace{0.cm}\includegraphics[trim={1cm 0cm 0cm 0cm},clip,width=0.45\textwidth,clip=true]{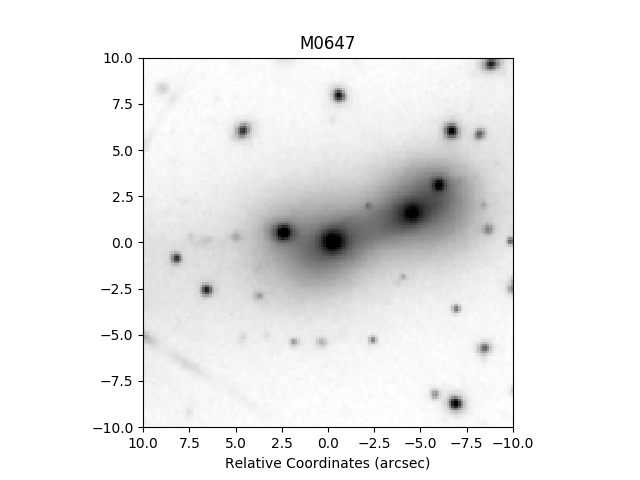}}\\ 
\caption{Archival $20^{\prime\prime}\times20^{\prime\prime}$ {\it HST} images for the BCGs with tentative CO detections and M0647, which has a massive western companion. The images are  taken with the infrared F160W filter on-board the WFC3 camera of {\it HST} and are centered at the BCG coordinates. North is up, East is left.}\label{fig:HSTimages}
\end{figure*}

\begin{appendix}
\onecolumn{
\section{CO spectra and \textit{HST} images}\label{app:appendix}
Here we report here the CO spectra and {\it HST} images for the CLASH BCGs in our sample with tentative CO detections, namely, M0329, A1423, M1206, and M2129. 
In Fig.~\ref{fig:HSTimages} we also report the archival {\it HST} image of the M0647 BCG which shows a massive Western companion.  
}
\begin{figure*}[h!]\centering
\captionsetup[subfigure]{labelformat=empty}
\subfloat[]{\hspace{0.cm}\includegraphics[trim={0cm 0cm 0cm 
0cm},clip,width=0.7\textwidth,clip=true]{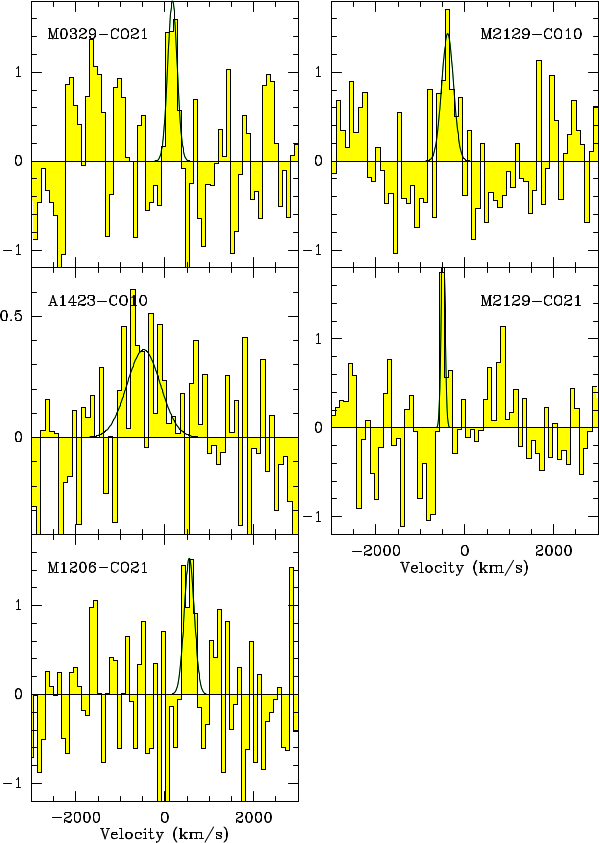}}
\caption{Baseline-subtracted spectra for the BCGs with tentative CO detections, obtained with the IRAM 30m. In all panels the solid line shows the Gaussian best fit. In each panel, the x-axis shows the relative velocity with respect to the BCG redshift, as reported in Table~\ref{tab:BCG_properties}, while in the y-axis Tmb is shown in units of mK.}\label{fig:hint_detection_spectra}
\end{figure*}

\end{appendix}


\begin{thebibliography}{}
\bibitem[Annunziatella et al.(2014)]{Annunziatella2014} Annunziatella, M., Biviano, A., Mercurio, A. et al. 2014, A\&A, 571, 80
\bibitem[Annunziatella et al.(2016)]{Annunziatella2016} Annunziatella, M., Mercurio, A., Biviano, A. et al. 2016, A\&A, 585, 160
\bibitem[Aravena et al.(2012)]{Aravena2012} Aravena, M., Carilli, C.~L., Salvato, M. et al. 2012, MNRAS, 426, 258 
\bibitem[Bezanson et al.(2019)]{Bezanson2019} { Bezanson, R., Spilker, J.,  Williams, C.~C. et al. 2019, ApJ, 873, 19}
\bibitem[Bigiel et al.(2008)]{Bigiel2008} Bigiel F., Leroy A., Walter F. et al. 2008, AJ, 136, 2846
\bibitem[Bolatto et al.(2013)]{Bolatto2013}  Bolatto, A.~T., Wolfire, M., \& Leroy, A.~K 2013, ARA\&A, 51, 207
\bibitem[Bonaventura et al.(2017)]{Bonaventura2017} Bonaventura, N.~R., Webb, T.~M.~A.,  Muzzin, A. et al. 2017, MNRAS, 469, 1259
\bibitem[Bothwell et al.(2013)]{Bothwell2013} Bothwell, M. S., Smail, I., Chapman, S. C., et al. 2013, MNRAS, 429, 3047
\bibitem[Burke et al.(2015)]{Burke2015} Burke, C., Hilton, M., Collins, C. et al. 2015, 2015, MNRAS, 449, 2353
\bibitem[Caminha et al.(2019)]{Caminha2019} Caminha, G.~B., Rosati, P., Grillo, C. et al. 2019, A\&A, 632, 36
\bibitem[Carilli \& Walter(2013)]{Carilli_Walter2013} Carilli, C.~L., \& Walter, F. 2013, ARA\&A, 51, 105
\bibitem[Castignani et al.(2018)]{Castignani2018} Castignani, G., Combes, F., Salom\'{e}, P. et al. 2018,  A\&A, 617, 103
\bibitem[Castignani et al.(2019)]{Castignani2019} Castignani, G., Combes, F., Salom\'{e}, P. et al. 2019, A\&A, 623, 48
\bibitem[Castignani et al.(2020a)]{Castignani2020a} Castignani, G., Combes, F., Salom\'{e}, P. et al. 2020a, A\&A, 635, 32 
\bibitem[Castignani et al.(2020b)]{Castignani2020b} Castignani, G., Combes, F., \& Salom\'{e}, P., 2020b,  A\&A 635, L10 
\bibitem[Castignani et al.(2020c)]{Castignani2020c} Castignani, G., Jablonka, P., Combes, F. et al. 2020c, arXiv:200203818
\bibitem[Chabrier(2003)]{Chabrier2003} Chabrier G., 2003, PASP, 115, 763
\bibitem[Coogan et al.(2018)]{Coogan2018} { Coogan, R.~T., Daddi, E., Sargent, M.~T. et al. 2018, MNRAS, 479, 703}
\bibitem[Cooke et al.(2016)]{Cooke2016} Cooke, K.~C., O'Dea, C.~P., Baum, S.~A. et al. 2016, ApJ, 833, 224
\bibitem[Cybulski et al.(2016)]{Cybulski2016} Cybulski, R., Yun, M.~S., Erickson, N. et al. 2016, MNRAS, 459, 3287
\bibitem[Daddi et al.(2015)]{Daddi2015} Daddi, E., Dannerbauer, H., Liu, D. et al. 2015, A\&A, 577, 46
\bibitem[DeMaio et al.(2019)]{DeMaio2019} DeMaio, T., Gonzalez, A.~H., Zabludoff, A. et al. 2020, MNRAS, 491, 3751
\bibitem[Devereux et al.(1994)]{Devereux1994} Devereux, N., Taniguchi, Y., Sanders, D. B., et al. 1994, AJ, 107, 2006
\bibitem[Donahue et al.(2014)]{Donahue2014} Donahue, M., Voit, G. M., Mahdavi, A., et al. 2014, ApJ, 794, 136
\bibitem[Donahue et al.(2015)]{Donahue2015}Donahue, M., Connor, T., Fogarty, K., et al. 2015, ApJ, 805, 177
\bibitem[Donahue et al.(2016)]{Donahue2016} Donahue, M., Ettori, S., Rasia, E. et al. 2016, ApJ, 819, 36, 2016 
\bibitem[Durret et al.(2019)]{Durret2019} Durret, F., Tarricq, Y., M\'{a}rquez, I. et al. 2019, A\&A, 622, 78
\bibitem[Edge(2001)]{Edge2001} Edge, A. C., 2001, MNRAS, 328, 762
\bibitem[Fabian(1994)]{Fabian1994} Fabian, A.~C. 1994, ARA\&A,32, 277
\bibitem[Fogarty et al.(2015)]{Fogarty2015} Fogarty, K., Postman, M., Connor, T. et al. 2015, ApJ, 813, 117
\bibitem[Fogarty et al.(2017)]{Fogarty2017} Fogarty, K., Postman, M., Larson, R. et al. 2017, ApJ, 846, 103
\bibitem[Fogarty et al.(2019)]{Fogarty2019} Fogarty, K., Postman, M., Li, Y. et al. 2019, ApJ, 879, 103
\bibitem[Fraser-McKelvie et al.(2014)]{FraserMcKelvie2014} Fraser-McKelvie, A., Brown, M.~J.~I., \& Pimbblet, K.~A., 2014, MNRAS, 444, 63
\bibitem[Freundlich et al.(2019)]{Freundlich2019} Freundlich, J., Combes, F., Tacconi, L. J. et al., 2019, A\&A, 622, 105
\bibitem[Gaspari et al.(2012)]{Gaspari2012} Gaspari, M., Ruszkowski, M., \& Sharma, P. 2012, ApJ, 746, 94
\bibitem[Geach et al.(2011)]{Geach2011} Geach, J.~E., Smail, I., Moran, S.~M. et al. 2011, ApJL, 730, 19
\bibitem[Gobat et al.(2018)]{Gobat2018} { Gobat, R., Daddi, E., Magdis, G., et al. 2018, Nature Astronomy, 2, 239}
\bibitem[Hamer et al.(2012)]{Hamer2012} Hamer, S.~L.,  Edge, A.~C., Swinbank, A.~M. et al. 2012, MNRAS, 421, 3409 
\bibitem[Hausman \& Ostriker(1978)]{Hausman_Ostriker1978} Hausman, M.~A. \& Ostriker, J.~P., 1978. , ApJ, 224, 320
\bibitem[Hayashi et al.(2018)]{Hayashi2018} { Hayashi, M., Tadaki, K., Kodama, T. et al. 2018, ApJ, 856, 118}
\bibitem[Hicks et al.(2010)]{Hicks2010} Hicks, A.~K., Mushotzky, R., \& Donahue, M. 2010, ApJ,719, 1844
\bibitem[Jablonka et al.(2013)]{Jablonka2013}  Jablonka, P., Combes, F., Rines, K. et al. 2013, A\&A, 557, 103
\bibitem[Kneissl et al.(2019)]{Kneissl2019} { Kneissl, R., del Carmen Polletta, M., Martinache, C. et al. 2019, A\&A, 625, 96}
\bibitem[Kramer et al.(2013)]{Kramer2013}  Kramer, C.,  Pe\~{n}alver, J. \& Greve A., 2013, \emph{Improvement of the IRAM 30m telescope beam pattern}, www.iram-institute.org$/$medias$/$uploads$/$eb2013-v8.2.pdf
\bibitem[Lauer et al.(2014)]{Lauer2014} Lauer, T.~R., Postman, M., Strauss, M.~A. et al., 2014, ApJ, 797, 82
\bibitem[Lee et al.(2017)]{Lee2017}  Lee, M.~M., Tanaka, I., Kawabe, R. et al. 2017,  ApJ, 842, 55
\bibitem[Leroy et al.(2013)]{Leroy2013} Leroy A.~K., Walter F., Sandstrom K. et al. 2013, AJ, 146, 19
\bibitem[Li \& Bryan(2014)]{Li_Bryan2014} Li, Y., \& Bryan, G. L. 2014, ApJ, 789, 153
\bibitem[Li et al.(2015)]{Li2015} Li, Y., Bryan, G. L., Ruszkowski, M., et al. 2015, ApJ, 811, 72
\bibitem[Madau \& Dickinson(2014)]{Madau_Dickinson2014} Madau, P. \& Dickinson, M., 2014, ARA\&A, 52, 415
\bibitem[McDonald et al.(2013)]{McDonald2013} McDonald, M., Benson, B., Veilleux, S. et al. 2013, ApJL, 765, 37
\bibitem[McDonald et al.(2014)]{McDonald2014} McDonald, M., Swinbank, M., Edge, A.~C. et al. 2014, ApJ, 784, 18
\bibitem[McNamara et al.(2000)]{McNamara2000} McNamara, B.~R., Wise, M., Nulsen, P. E. J., et al. 2000, ApJL, 534, 135
\bibitem[McNamara \& Nulsen(2012)]{McNamara_Nulsen2012} McNamara, B.~R., \& Nulsen, P.~E.~J. 2012, NJPh, 14, 055023
\bibitem[McNamara et al.(2014)]{McNamara2014} McNamara, B.~R., Russell, H.~R., Nulsen, P.~E.~J. et al. 2014, ApJ, 785, 44 
\bibitem[Noble et al.(2017)]{Noble2017} Noble, A.~G., McDonald, M., Muzzin, A., et al. 2017, ApJ, 842, 21
\bibitem[Noble et al.(2019)]{Noble2019} { Noble, A.~G., Muzzin, A., McDonald, M. et al. 2019, ApJ, 870, 56}
\bibitem[Olivares et al.(2019)]{Olivares2019}  Olivares, V., Salom\'{e}, P., Combes, F., et al. 2019, A\&A, 631, 22O 
\bibitem[Papadopoulos et al.(2000)]{Papadopoulos2000} Papadopoulos, P.~P., R\"{o}ttgering, H.~J.~A., van der Werf, P.~P. et al. 2000, ApJ, 528, 626
\bibitem[Papadopoulos et al.(2012)]{Papadopoulos2012} Papadopoulos, P.~P., van der Werf, P.~P., Xilouris, E.~M. et al. 2012, MNRAS, 426, 2601
\bibitem[Peng et al.(2002)]{Peng2002} {Peng, C.~Y., Ho, L.~C., Impey, C.~D. et al. 2002, AJ, 124, 266}
\bibitem[Peng et al.(2010)]{Peng2010} { Peng, C.~Y., Ho, L.~C., Impey, C.~D. et al. 2010, AJ, 139, 2097}
\bibitem[Peterson et al.(2003)]{Peterson2003} Peterson, J.~R., Kahn, S.~M., Paerels, F.~B.~S., et al. 2003, ApJ, 590, 207
\bibitem[Peterson \& Fabian(2006)]{Peterson_Fabian2006} Peterson, J.~R., \& Fabian, A.~C. 2006, PhR, 427, 1
\bibitem[Planck Collaboration(2018)]{PlanckCollaborationVI2018} Planck Collaboration results VI, 2018, arXiv:180706209
\bibitem[Postman et al.(2012a)]{Postman2012a} Postman, M., Coe, D., Ben\'{i}tez, N., et al. 2012a, ApJS, 199, 25
\bibitem[Postman et al.(2012b)]{Postman2012b} Postman, M., Lauer, T.~R., Donahue, M., et al. 2012b, ApJ, 756, 159 
\bibitem[Rawle et al.(2012)]{Rawle2012} Rawle, T.~D., Edge, A.~C., Egami, E., et al. 2012, ApJ, 747, 29
\bibitem[Riechers et al.(2011)]{Riechers2011} Riechers D.~A., Carilli C.~L., Maddalena R.~J et al. 2011, ApJL, 739, 32
\bibitem[Riess(2019)]{Riess2019b} Riess, A.~G., 2019 Nature Reviews Physics, 2, 10
\bibitem[Rudnick et al.(2017)]{Rudnick2017} Rudnick, G., Hodge, J., Walter, F. et al. 2017, ApJ, 849, 27
\bibitem[Russell et al.(2014)]{Russell2014} Russell, H.~R., McNamara, B.~R.,  Edge, A.~C. et al. 2014,  ApJ, 784, 78 
\bibitem[Russell et al.(2019)]{Russell2019} Russell, H.~R., McNamara, B.~R., Fabian, A.~C. et al. 2019, MNRAS, 490, 3025
\bibitem[Salom\'{e} \& Combes(2003)]{Salome_Combes2003} Salom\'{e} P. \&  Combes, F. 2003, A\&A, 412, 657
\bibitem[Salom\'{e} et al.(2006)]{Salome2006} Salom\'{e} P., Combes, F., Edge, A.~C. et al. 2006, A\&A, 454, 437
\bibitem[Salpeter(1955)]{Salpeter1955} Salpeter, E.~E. 1955,ApJ,121, 161
\bibitem[Sargent et al.(2015)]{Sargent2015} Sargent, M. T., Daddi, E., Bournaud, F., et al. 2015, ApJL, 806, L20
\bibitem[Schruba et al.(2011)]{Schruba2011} Schruba A., Leroy A.~K., Walter F. et al. 2011, AJ, 142, 37  
\bibitem[Solomon et al.(1997)]{Solomon1997} { Solomon, P.~M., Downes, D., Radford, S.~J.~E. et al. 1997, ApJ, 478, 144}
\bibitem[Solomon \& Vanden Bout(2005)]{Solomon_VandenBout2005} Solomon, P.~M. \& Vanden Bout, P.~A., 2005, ARA\&A, 43, 677
\bibitem[Speagle et al.(2014)]{Speagle2014} { Speagle, J.~S., Steinhardt, C.~L., Capak, P.~L. et al. 2014, ApJS, 214, 15}
\bibitem[Stott et al.(2012)]{Stott2012} Stott, J.~P., Hickox, R.~C., Edge, A.~C. et al., 2012, MNRAS, 422, 2213
\bibitem[Strazzullo et al.(2018)]{Strazzullo2018} Strazzullo, V., Coogan, R.~T., Daddi, E. et al. 2018, ApJ, 862, 64
\bibitem[Tacconi et al.(2018)]{Tacconi2018} Tacconi, L.~J., Genzel, R., Saintonge, A., et al. 2018, ApJ, 853, 179
\bibitem[Tadaki et al.(2019)]{Tadaki2019} Tadaki, K., Kodama, T., Hayashi, M. et al. 2019, PASJ, 71, 40
\bibitem[Tremblay et al.(2016)]{Tremblay2016} Tremblay, G.~R., Oonk, J.~B.~R., Combes, F. et al. 2016, Nature, 534, 218
\bibitem[Trudeau et al.(2019)]{Trudeau2019} Trudeau, A., Webb, T., Hlavacek-Larrondo, J. et al. 2019, MNRAS, 487, 1210
\bibitem[van der Wel et al.(2014)]{vanderWel2014} { van der Wel, A., Franx, M., van Dokkum, P.~G. et al. 2014, ApJ, 788, 28}
\bibitem[Voit(2005)]{Voit2005} Voit, G. Mark, 2005, RvMP, 77, 207
\bibitem[Voit \& Donahue(2015)]{Voit_Donahue2015} Voit, G.~M., \& Donahue, M. 2015, ApJL, 799, L1
\bibitem[Voit et al.(2017)]{Voit2017} Voit, G.~M., Meece, G., Li, Y., et al. 2017, ApJ, 845, 80
\bibitem[Wagg et al.(2012)]{Wagg2012} Wagg, J., Pope, A., Alberts, S. et al. 2012, ApJ, 752, 91 
\bibitem[Webb et al.(2015)]{Webb2015} Webb, T.~M.~A., Muzzin, A., Noble, A. et al. 2015, ApJ, 814, 96
\bibitem[Webb et al.(2017)]{Webb2017} Webb, T.~M.~A., Lowenthal, J., Yun, M. et al. 2017, ApJ, 844, 17
\bibitem[White(1976)]{White1976} White, S.~D.~M., 1976, MNRAS, 177, 717
\bibitem[Young et al.(1995)]{Young1995} Young, J.~S., Xie, S., Tacconi, L. et al. 1995, ApJS, 98, 219
\bibitem[Yu et al.(2018)]{Yu2018} Yu, H., Tozzi, P., van Weeren, R. et al. 2018, ApJ, 853, 100
\bibitem[Zirbel(1996)]{Zirbel1996} Zirbel, E.~L. 1996, ApJ, 473, 713
\end{thebibliography}
\end{document}